\documentclass[12pt,preprint]{aastex}

\newcommand{\chandra}{{\em{Chandra}}}
\newcommand{\cxo}{{\it{Chandra X-ray Observatory}}}
\newcommand{\ty}{Tycho}
\newcommand{\einstein}{{\em{Einstein}}}
\newcommand{\ginga}{{\em{Ginga}}}

\shorttitle{Cosmic Ray Acceleration in Tycho's SNR}
\shortauthors{Warren et al.}

\begin{document}

\title{Cosmic Ray Acceleration at the Forward Shock in Tycho's
Supernova Remnant: Evidence from \chandra\ X-ray Observations}

\author{Jessica S. Warren,\altaffilmark{1}
 John P. Hughes,\altaffilmark{1}
 Carles Badenes,\altaffilmark{1}
 Parviz Ghavamian,\altaffilmark{2}
 Christopher F. McKee,\altaffilmark{3}
 David Moffett,\altaffilmark{4}
 Paul P. Plucinsky,\altaffilmark{5}
 Cara Rakowski,\altaffilmark{5}
 Estela Reynoso,\altaffilmark{6}
and Patrick Slane\altaffilmark{5}
}

\altaffiltext{1}{Department of Physics \& Astronomy, Rutgers University, 136 Frelinghuysen Road, Piscataway, NJ 08854-8019; jesawyer@physics.rutgers.edu, jph@physics.rutgers.edu}
\altaffiltext{2}{Department of Physics \& Astronomy, Johns Hopkins University, 3400 N Charles Street, Baltimore, MD 21218-2686}
\altaffiltext{3}{Departments of Physics \& Astronomy, University of California, Berkeley, CA 94720}
\altaffiltext{4}{Physics Department, Furman University, 3300 Poinsett Highway, Greenville, SC 29613}
\altaffiltext{5}{Harvard-Smithsonian Center for Astrophysics, 60 Garden Street, Cambridge, MA 02138}
\altaffiltext{6}{Instituto de Astronom\'{\i}a y F\'{\i}sica del Espacio, C.C. 67, Sucursal 28, 1428 Buenos Aires, Argentina}

\begin{abstract}

We present evidence for cosmic ray acceleration at the forward shock
in \ty's supernova remnant (SNR) from three X-ray observables: (1) the
proximity of the contact discontinuity to the forward shock, or blast
wave, (2) the morphology of the emission from the rim of \ty, and (3)
the spectral nature of the rim emission.  We determine the locations
of the blast wave (BW), contact discontinuity (CD), and reverse shock
(RS) around the rim of \ty's supernova remnant using a principal
component analysis and other methods applied to new \chandra\ data.
The azimuthal-angle-averaged radius of the BW is 251$\arcsec$.  For
the CD and RS we find  average radii of 241$\arcsec$ and 183$\arcsec$,
respectively.  Taking account of projection effects, we find ratios of
1~:~0.93~:~0.70 (BW~:~CD~:~RS).  We show these values to be
inconsistent with adiabatic hydrodynamical models of SNR evolution.
The CD:BW ratio can be explained if cosmic ray acceleration of {\em
ions} is occurring at the forward shock.  The RS:BW ratio, as well as
the strong Fe K$\alpha$ emission from the \ty\ ejecta, imply that the
RS is not accelerating cosmic rays.  We also extract radial profiles
from $\sim$34\% of the rim of \ty\ and compare them to models of
surface brightness profiles behind the BW for a purely thermal plasma
with an adiabatic shock.  The observed morphology of the rim is much
more strongly peaked than predicted by the model, indicating that such
thermal emission is implausible here.  Spectral analysis also implies
that the rim emission is non-thermal in nature, lending further
support to the idea that \ty's forward shock is accelerating cosmic
rays.

\end{abstract}


\section{Introduction}

With the clear detection of non-thermal X-ray synchrotron emission
from the rims of the remnant of SN 1006 by \citet{koy95}, it was
established that the shocks in supernova remnants (SNRs) can
accelerate cosmic ray electrons to TeV energies.  These energies are
close to, but do not quite reach, the ``knee'' in the cosmic ray
spectrum around 300 TeV.  Since the work on SN 1006, several other
SNRs have been found to emit X-ray synchrotron radiation, including
Cas A \citep{allen97}, G266.2-1.2 \citep{slane}, and G347.3-0.5
\citep{sla99}, implying cosmic ray shock acceleration may be common
in young SNRs.

Still, much remains unknown or observationally unsubstantiated.  How
efficient is the acceleration mechanism?  The amount of energy the
shock puts into accelerating particles should be large \citep{dec00},
but independent observational constraints on this are limited
\citep[e.g.,][]{hrd00}.  Are cosmic ray ions being accelerated?  X-ray
synchrotron radiation is evidence of relativistic cosmic ray
electrons.  Direct evidence of the presence of cosmic ray protons
comes in the form of $\gamma$-rays via pion-decay \citep{el05}.  This
has not yet been conclusively detected, partly because such a
signature is difficult to extract from the observations
\citep[see][and references therein]{baring,malkov}.  How does cosmic
ray acceleration affect the dynamics of an SNR?  The dynamical
evolution of an SNR depends on the properties (e.g., explosion energy,
mass of ejecta, ambient density) of its supernova (SN) and surrounding
environment.  These factors influence the rate at which the forward
shock expands into the interstellar medium (ISM), where the contact
discontinuity (which separates the SN ejecta from the swept-up ISM)
lies, and the rate at which the reverse shock propagates into the
ejecta.  If the shocks are accelerating cosmic rays, particularly
ions, their rates of propagation will be affected \citep{el05}.

In the 0.2-10 keV X-ray band where superb imaging of SNRs is possible
using \chandra, the key to addressing these questions lies in cleanly
separating regions whose emission is dominated by featureless spectra
from other regions that are line dominated.  In all but the most
exceptional cases, one can identify the dominant emission processes in
these two types of regions to be X-ray synchrotron radiation or
thermal radiation from shock-heated ejecta or ISM, respectively.  With
the power to discriminate where these different processes occur in an
SNR, one can identify the sites where electrons are being accelerated
to high energies and where the shocked material (both ejecta and
ambient) is situated. With knowledge of the locations of the blast
wave and contact discontinuity one constrains the dynamical state of
the remnant and can then investigate what role, if any, cosmic ray
acceleration plays.

In this article, we demonstrate the separation of
featureless-dominated and line-dominated emission in the remnant of SN
1572 (hereafter \ty) using the results of a principal component
analysis (PCA).  PCA is a mathematical operation that reduces the
dimensionality of a dataset.  This is accomplished by finding new
variables to characterize the data that are linear combinations of the
original variables.  These new variables, or principal components, are
chosen such that they maximize the variance of the data.  For \ty, our
data consist of the spectra of thousands of spatial regions in the
SNR.  Our original variables are spectral channels, or energy bands.

The analysis in our paper is the first application of PCA to data from
an SNR.  \ty, however, has been studied extensively with other means.
It is widely believed to be the remnant of a Type Ia supernova (SN
Ia).  Tycho's forward shock is filamentary and shows featureless X-ray
emission in the \chandra\ band, of which some fraction is likely to be
non-thermal \citep{hwa02}.  Optical observations reveal thin
Balmer-dominated filaments in the north and east of the remnant
\citep{smith,ghav}.  In the radio, \ty\ shows a limb-brightened shell
and a thicker, inner shell \citep{dic91}.  These authors and, more
recently, \citet{ks00} found that the radio emission is consistent
with the acceleration of cosmic ray electrons by the forward shock.
\citet{re92} specifically modeled the non-linear first-order Fermi
acceleration mechanism (also called diffusive shock acceleration,
DSA), which is believed to be the process by which the shock
accelerates cosmic rays.  They found good agreement between the model
and the radio spectrum of \ty.  \citet{el01} compared a model of DSA
to \ty's combined radio and X-ray spectrum, including limits from
$\gamma$-ray observations.  He found that the shock can accelerate
particles up to TeV energies, whose synchrotron signature should be
detected in the X-ray band.  Additionally, hard X-rays out to $\sim$20
keV have been detected in \ty\ with the \ginga\ satellite
\citep{fink94}.  The spectrum at these energies cannot be fit with a
single bremsstrahlung component, but requires a second, ``flat''
component to fit the emission.

The ejecta of \ty, only visible in the X-rays, are spread throughout
the interior in a clumpy shell-like distribution and are mainly Si-
and Fe-rich \citep{hwa98,dec01}.  The Fe emission in Tycho appears to
be stratified; the Fe L emission peaks at a larger radius than the Fe
K emission \citep{hg97}.  A hot Fe component is required to reproduce
the flux in the Fe K$\alpha$ line in the integrated spectrum, implying
there is a temperature gradient in the ejecta \citep{hwa98}.

The distance to \ty\ is estimated to be between 1.5--3 kpc
\citep{smith}, depending on the method used.  Our main conclusions are
based on ratios of observed radii and are therefore distance
independent. When quoting results in physical units, we explicitly
state the distance value used (2.3 kpc).

The plan of our paper is as follows.  In \S 2 we describe our
observations and the PCA technique.  The results of the PCA allow us
to locate the contact discontinuity, which we do in \S 3 along with
locating the blast wave.  In \S 4 we examine the azimuthal variation
of radial fluctuations of the contact discontinuity via a power
spectrum analysis.  We examine the radial morphology of the rim in \S
5.  To round out our discussion of the fluid discontinuities in \ty,
in \S 6 we determine the location of the reverse shock.  Finally, we
discuss our results in \S 7 and summarize the paper in \S 8.

\section{Observations and Techniques}

We observed \ty\ with the \chandra\ Advanced CCD Imaging Spectrometer
(ACIS-I) detector on 2003 April 29-30 for $\sim$150 ks (ObsID 3837) in
timed exposure mode.  All four chips were active, the aimpoint was
located at the SNR's geometric center, and the full 8$\arcmin$
diameter extent of \ty\ was imaged. Unfortunately, the instrument was
in dither mode (which ensures that no single portion of the image
falls on a bad pixel for the entire observation) for only the latter
68 ks of the exposure.  However, the lack of dithering for roughly
half the observation had no significant impact on our analysis.  We
applied standard reduction techniques to the events, filtering on
grade (retaining the usual values 02346) and bad pixels, and
eliminating times of high background.  We removed pixel randomization
and applied the time-dependent gain correction on an event-by-event
basis.  After all processing, the final exposure time was 145.2 ks.

\subsection{Three-color Image}

Figure~\ref{color} (top left) is a composite of three X-ray energy
bands that are dominated by emission from Fe L-shell (red), Si K-shell
(green), and high energy (4--6 keV) continuum (blue).  The blue
component is spectrally hard and featureless.  It predominates at the
outer rim in the form of thin, moderately high surface brightness
filaments that can be traced continuously for nearly 180$\degr$ around
the rim from the north to southwest.  This emission also spreads
throughout the interior. Two features in particular stand out: a small
arc in the southeast, and a larger, fainter arc that runs northward
from the southeast.  These are possibly more of the filamentary
structures at the rim that are just seen in projection.  There are
also two more knot-like structures in this energy range in the
southwest, slightly behind the rim \citep[one of which was noted
by][]{hwa02}.  The ``fleecy'' green component is rich in Si and S and
generally fills the interior.  At several locations it extends to the
edge of the SNR, breaking through the blue rim emission.  The red
component is more Fe-rich and appears particularly enhanced in a small
region on the eastern rim \citep[first noted by][]{van95} and a more
extensive region interior to the Si-rich zone in the northwest.  Both
the red and green components are more knot-like in structure, as
opposed to the filamentary nature of the blue component.

\subsection{PCA Technique}

Already we see from the 3-color image that emission from the rim, or
forward shock, stands out in the 4--6 keV continuum range, while the
interior emission is line-dominated. It seems likely that judicious
use of the 3-color image or some combination of ratio images would
allow us to separate the forward shock (ISM) emission from the
interior (ejecta) emission. Similar problems have presented themselves
in many of our previous studies with \chandra, and so rather than
pursue a limited {\it ad hoc} approach here just for \ty, we decided
to investigate principal component analysis (PCA), which is a general
mathematical technique that is quantitative, rigorous, and less
subject to observer bias.

We present here a brief summary of our application of PCA to \ty; an
in-depth account and further motivation will be given elsewhere
\citep{wh05}. A mathematical description of PCA can be found in
\citet{mh}.  We extracted, from small regions across the entire
remnant, a large number of spectra in 12 broad spectral channels that
were chosen to isolate important spectral features (e.g., the three
energy bands in the color composite, see Table 1). These spectra can
be considered as points in a 12-dimensional space, whose axes
correspond to the counts in each spectral channel.  PCA is a
mathematical operation that identifies a new set of 12 axes along
which the variance of the data points is maximized.  These new axes,
called principal components, are a rank-ordered set of spectral
templates that quantify both the amount and type of spectral variation
within the data.  In other words, the principal components are linear
combinations of the original 12 spectral channels.  The top right
image in Figure~\ref{pc1} represents one linear combination of images
made in those original 12 spectral bands.  The weight factors
quantifying the contribution from each individual band image are
determined from the entire dataset through the well-defined
mathematical process of PCA.  The image is conceptually analogous to
the hardness ratio images that many X-ray astronomers use to identify
spectral variations with position.  The PCA is a significant
generalization of that technique, since it uses the entire spectral
range of a dataset.

We found that the Tycho spectral data points, when projected onto the
new principal axes, were distributed in a well-behaved manner: the
points were spread out along, and very nearly parallel to, the
principal axes. The spread of data points along one of the new
principal axes need not correspond uniquely to variations in a single
physical parameter.  In our application here, however, we found that
the first principal axis, which accounts for $\sim$43\% of the total
spectral variation, allows for a simple physical interpretation. The
top right panel of Figure~\ref{pc1} shows the projection of the
\chandra\ data onto this first principal axis (hereafter PC1).
Comparison with the color image suggests that regions appearing
light-colored in the PC1 image correspond to the ``fleecy'' Si- and
Fe-rich emission, while the dark regions correspond to the hard
continuum emission.

\subsection{Spectra}

To further support our interpretation of PC1 we extracted and studied
the spectra from a large number of regions in \ty\ (in the following
we present and discuss only a representative subset of these spectra).
The spectra from three regions along the N, S, and SW rim,
corresponding to extreme values of PC1, are plotted in
Figure~\ref{spectra} (top panel).  They were well-fitted by absorbed
power-law spectral models ($\chi^2_{reduced} \sim 0.8$) with parameter
values in the range $N_{\rm H} = (5.8-6.9)\times 10^{21}$ cm$^{-2}$
and $\alpha_P = 2.67-2.71$.  Pure bremsstrahlung models result in
equally good fits with parameter value ranges of $N_{\rm H} =
(3.9-4.8)\times 10^{21}$ cm$^{-2}$ and $kT = 2.1-2.2$ keV.  We discuss
the nature of the emission mechanism for the rim in \S
\ref{elacc} below.  Here, our intent is simply to demonstrate that
extreme values to one direction along the PC1 axis (i.e., the dark
portions of the PC1 image) correspond to featureless-dominated spectra
that are mainly distributed around the rim of \ty.

We extracted spectra from the reddish and yellowish regions in the
southeast in Figure~\ref{color} (top left), which also correspond to
light portions of the PC1 image.  These are plotted in
Figure~\ref{spectra} (bottom panel) and clearly show that the red region
is Fe-rich and the yellow region is Si-rich.  Although we do not show
them here \citep[a more detailed analysis of the ejecta is left
to][]{hw05}, we extracted several more spectra from various ``knots''
around the remnant that appear light-colored in the PC1 image.  All of
these are line-dominated, mainly by Si and S K$\alpha$.  It is thus
clear that the light portions of the PC1 image correspond to
metal-rich thermal spectra.  The PCA technique therefore has given us
a clear delineation between the two main spectral types in \ty: the
featureless rim emission and the metal-rich thermal emission in the
remnant interior.  From our PCA technique, we note as an important
addition that thermal emission from material with normal solar-type
composition is not a major spectral type in \ty.

Based on its morphology, the featureless component must be emission
from the blast wave (BW).  As we move in from the rim, Si and S line
emission becomes more prominent.  We interpret this thermal emission
appearing behind the rim as shocked ejecta, rather than the
progression of the ionization state behind the BW
\citep[e.g.,][]{hwa02}.  Moving in from the rim, the ejecta become
less diluted with the surrounding shocked ambient medium, simply due
to geometric projection.  The ejecta shell is clumpy and patchy.  In
the interior there is a correlation between the X-ray intensity
(Figure~\ref{color} top left) and PC1 image: where the overall
brightness is low, the spectral character shifts toward the
featureless type and vice versa. This is likely the effect of the
projection of the BW and ejecta shells that vary in thickness and
intrinsic intensity across the line of sight.  This interpretation
allows us to determine the location of the contact discontinuity (CD).

\section{Tycho's Blast Wave and Contact Discontinuity}\label{bwandcd}

We associate the maximum radial extent of the {\em broadband} X-ray
emission with the location of the BW, regardless of its spectral
character.  This ensures that the BW radius is always greater than or
equal to the CD radius.  To determine the position of the BW, we first
blocked our image to a scale of 0$\farcs$984 per pixel. At a given
azimuthal angle, we extracted the radial surface brightness profile
from the broadband image and determined the radius where the profile
first exceeded 8 cts pixel$^{-1}$ moving in from the outside (see the
top panel of Figure~\ref{radprof}).  We repeated this for angular steps
of 1$\degr$ around the entire SNR.  The surface brightness value used
(5.7$\times10^{-5}$ cts s$^{-1}$ arcsec$^{-2}$ or roughly 7 times the
background) provided a compromise in denoting the BW radius between
faint regions, as in the north and southeast, and bright regions, as
in the northwest and east.  At any given azimuthal angle, we expect
the accuracy in measuring the position of a specific contour level
(i.e., the location of the BW) to be of order the width of the
point-spread-function (PSF) divided by the square root of the number
of counts at that contour level. This evaluates to
$\sim$1$\arcsec$/$\sqrt 8 \sim 0\farcs3$ in our case.  The BW radius
as a function of azimuthal angle is plotted as the black curve in
Figure~\ref{radplot}.  The center for this determination
(R.A.=00$^\textrm{\scriptsize{h}}$25$^\textrm{\scriptsize{m}}$19$\fs$40,
decl.=+64$\degr$08$\arcmin$13$\farcs$98) was chosen to minimize the
ellipticity of the BW and is used as the center throughout.

We determined the location of the CD from the value of PC1 (i.e., the
level of ``darkness'' in the PC1 image) that discriminates between the
outermost ejecta and BW emission.  This was obtained from the
distribution of PC1 values, which has a well-defined peak (see
Figure~\ref{pc1dist}).  This peak comes from the large number of pixels
associated with a light yellow color in the PC1 image, which we
established as being thermal in nature.  The distribution around the
peak is well-fit by a Gaussian with $\sigma=0.028$ (the red curve in
Figure~\ref{pc1dist}), although there is an excess of PC1 values
corresponding to pixels in which the emission is dominated by the
featureless component.  We found the PC1 value needed to discriminate
between ejecta (thermal) and blast wave (featureless) emission by
first using the Gaussian to represent the thermal distribution {\em
alone}, then calculating the PC1 value lying 3-$\sigma$ away from the
peak on the side representing featureless emission.  Pixels with PC1
values lying below this 3-$\sigma$ threshold are unlikely to be
dominated by thermal emission.

To mark this position on the remnant, we took the intensity-weighted
average of PC1 values for regions of size 1$^\circ$ ($\sim$4$\arcsec$
at the rim) in azimuth by 3$\arcsec$ in radius. Moving in from the rim
in 1$\arcsec$ steps, we found the radius where the average PC1 value
first crossed the 3-$\sigma$ threshold.  We took this to be the radius
of the CD.  A contour representing this delineation is shown in the
PC1 image (Figure~\ref{pc1}), and a radial profile for a 1$\degr$-wide
wedge is shown in the middle panel of Figure~\ref{radprof}.  We also
looked at the contours corresponding to PC1 values 2-$\sigma$ and
4-$\sigma$ away from the peak and found no significant qualitative
differences.  We therefore use the 3-$\sigma$ criterion for
determining the CD in what follows, unless otherwise stated.

It is important to note that in determining the location of the CD
with the PCA technique we have utilized information from the {\em
entire} spectral range.  This is in contrast to other methods which
rely on a single line (e.g. Si K$\alpha$) to locate the CD.

We find the CD is close to the BW; the average radius of the CD
(241$\arcsec$) is 96\% that of the BW (251$\arcsec$).  If we use the
2-$\sigma$ and 4-$\sigma$ radii, we find the CD radius is 94\% and
98\% of the BW radius, respectively.  In Figure~\ref{radplot} we plot
the CD radius vs.~azimuthal angle.  One can see that the ejecta extend
out as far as the BW in discrete clumps at several azimuthal angles
(e.g., 40$\degr$--65$\degr$, 95$\degr$, 110$\degr$, 195$\degr$, and
360$\degr$).  These clumps are Si-rich, with the exception of the
Fe-rich knots between 105$\degr$ and 115$\degr$.  Figure~\ref{radplot}
also indicates that the CD is much more highly structured than the BW
on small angular scales.  We quantify this last statement by examining
the power spectrum of radial amplitude fluctuations.

\section{Power Spectrum}\label{powerspectrum}

The tabulated functions of radius vs.\ azimuthal angle for the BW and
CD were each decomposed in a harmonic sequence, resulting in
amplitudes and phase values for each of 180 trigonometric basis
functions used.  The power spectra (the square of the amplitudes) as a
function of wavenumber (defined as $k=n/R_{\rm ave}$, where $n$ is the
order of the trigonometric function and $R_{\rm ave}$ is the average
radius) are shown in Figure~\ref{fft} for the BW (black) and CD
(green).  The power peaks at low wavenumbers (long wavelengths) for
both the BW and CD and falls off at higher values. We find no evidence
for periodicity as suggested by \citet{vel98}.  Each power spectrum is well
described by a power law (as plotted on Figure~\ref{fft}); we find
${\rm P} \sim k^{-1.5}$ (CD) and ${\rm P} \sim k^{-2.2}$ (BW). The
single wavenumber with the largest power for the CD corresponds to
$n=6$ (or a wavelength in the azimuthal direction of 60$\degr$). The
amplitude of this component is 3.3\% of the average radius; the BW
shows nearly the same value. The minimum observed power in the BW
spectrum ($\sim$0.003 arcsec$^2$) corresponds to radial fluctuations
with an amplitude of $\sim$$0\farcs3$.  Since these arise from both
intrinsic and random (noise) variations in the measured position, we
conclude that the accuracy in the BW radial positions is on average
better than 0$\farcs$3, consistent with the uncertainty estimated
above.

The lack of a preferred scale length and the predominance of power at
low wavenumbers in the CD is qualitatively consistent with
expectations for structure driven by the Rayleigh-Taylor and
Kelvin-Helmholtz instabilities \citep[e.g.,][]{cbe92}.  At higher
wavenumbers the CD shows nearly an order of magnitude more power than
the BW, indicative of instabilities acting at the CD.  The level of
power at low wavenumbers is nearly the same for the BW as the CD, even
though the former is not subject to instability.  However, as the
direct observation of ejecta at the BW makes clear, some (and perhaps
most) of the BW's low order power is a result of these ejecta
protrusions, although it is possible that additional power comes from
environmental factors.

It is our hope that, in the not-too-distant future, multi-dimensional
hydrodynamical calculations of SNR evolution will have the resolution
to be able to produce radial power spectra similar to those we present
here from the observations.  Comparison of observed and simulated
power spectra may yield insights into the fluid properties of the
ejecta.

\section{Radial Morphology at the Rim of Tycho's SNR}

In the previous sections, we investigated the global structure of
\ty's rim as a function of azimuthal angle.  In order to gain more
insight into the precise nature and origin of this emission, here we
examine in detail the radial morphology of the rim. We began with the
4--6 keV band image, which is largely uncontaminated by line emission.
In selecting regions we avoided parts of the rim that were
contaminated by ejecta (identified using our 3-$\sigma$ PC1 criterion)
or where the edge was clearly structured (i.e., from multiple
overlapping rims projected on top of each other). This left us with
seven separate azimuthal regions covering $\sim$34\% of the entire rim
(see Figure~\ref{contimg}, bottom left). Rather than extract a typical
radial profile by summing over a range of azimuthal angles for a fixed
set of radial bins, we accounted for the azimuthal rippling of the BW
location within a given extraction region. First, the azimuthal angle
of each pixel in the extraction region was determined. With this angle
it was possible to look up the location of the BW in radius (i.e.,
from Figure~\ref{radplot}) and determine the radial distance of the
pixel behind (or ahead of) the BW.  The number of detected X-ray
events in the pixel was then added to a radial profile binned with
respect to the BW location.  A radial profile of the exposure map
(produced using standard CIAO software) was made in the same way and
then divided into the count rate profile.  The resulting surface
brightness profiles for the seven regions are plotted as the crosses
in Figure~\ref{rimprof}. Note that each profile in the figure has been
plotted so that the mean BW radius for the azimuthal region is
approximately correct. In all cases, while moving inward, the profiles
rise rapidly from the background level to a peak and then fall-off in
brightness further in toward the interior.  The widths of the rim as
seen here in projection are quite narrow, of order only a few
arcseconds, as previously noted by \citet{hwa02} and \citet{bam}.

\subsection{Rim Widths from Spherical Shell Models \label{rim}}

Here we consider two simple models for the intrinsic three-dimensional
structure of the rims.  Both cases assume the X-ray emission comes
from a geometrically thin spherical shell.  In the first case the
emissivity is considered to be uniform within the shell; the other
model, following \citet{bv04}, assumes the emissivity falls off
exponentially behind the shock, which is defined to be the outer
radius of the shell.  The relevant parameters are the outer radius of
the shell, the shell thickness (or the 1/e length of the exponential
profile), an intensity scale factor, and a background level (assumed
spatially uniform).  The models are projected to the plane of the sky
and convolved with the \chandra\ point-spread-function (PSF)
appropriate to each region. PSF simulations were generated using the
\chandra\ Ray Tracer (ChaRT)\footnote{ChaRT and MARX are available at:
http://asc.harvard.edu/chart/index.html} for a photon energy of 4.8
keV (the mean weighted energy in the 4--6 keV band).  The input
location to ChaRT was taken as the center of each azimuthal region at
the BW. The FWHM radial widths of the PSF profiles varied from
1$^{\prime\prime}$ to 2$^{\prime\prime}$.  Our fits improved
substantially when convolution with the PSF was included (especially
for the exponential profile), clearly demonstrating the value and
importance of including this in the fitting process.

We plot the best-fit uniform-emissivity projected shell models
convolved with the PSF as the solid histograms in Figure~\ref{rimprof}.
Nearly all of the rim profiles are well-described by this simple
model, although in two cases (regions covering azimuthal angles
312$^\circ$--318$^\circ$ and 273$^\circ$--289$^\circ$) the observed
surface brightness profiles are more limb-brightened than the model.
This cannot be solved by merely reducing the intrinsic thickness of
the shell model since this causes the width of the projected profile
to decrease, thereby degrading the fit. We surmise that these
azimuthal locations may be places where our assumption of a
spherically symmetric shell breaks down. The thickness of the shells
range from 2$\farcs$0--5$\farcs$7 for the uniform emissivity case.

For the case where the emissivity falls exponentially with distance
from the shock front we use the approximate formula\footnote{In
\citet{bv04} the correct relation appears as eqn.~6. There are
typographical errors in the same relation presented as eqn.~1 in
\citet{vbk05}.} given by \citet{bv04} for the projected profile.  The
quality of the fits here is just about as good as in the previous
case, although the best fit ``widths'' (i.e., 1/e lengths) are much
smaller and range over values of 0$\farcs$4--2$\farcs$9.

Since neither the uniform nor exponential profile cases is a clear
winner in terms of quality of fit and given the absence of any strong
astrophysical motivation to support one or the other, we will carry
both sets of width values in our subsequent discussions and
calculations.

\subsection{Rim Morphology for Shocked Plasma at the Blast 
Wave \label{morph}}

The extremely thin filaments around the outer rim of Tycho, first
noted by \citet{hwa02}, are not what we would expect from standard
adiabatic hydrodynamical evolutionary models.  Our expectation from
such models is for a much thicker region of thermal emission,
determined by the profiles of gas density, temperature, and ionization
conditions behind the BW. These thermodynamic quantities themselves
depend on the specific physical conditions and evolutionary state of
the remnant.  Although the gas density behind the BW is highest right
at the shock front, other thermodynamic quantities evolve with time
(i.e., distance behind the BW) and may not conspire to produce peak
emissivity near the rim.

To investigate this quantitatively we have produced surface brightness
profiles behind the BW for a purely thermal plasma with an adiabatic
shock.  We applied the hydrodynamic and ionization codes described in
\citet{bad03} to the shocked ambient medium (AM).  These assume the
unshocked AM has a constant density of $1\times10^{-24}$ g cm$^{-3}$
and solar composition, and we varied the amount of collisionless
electron heating at the BW.  These calculations are very similar to
those described in \citet{bor01} - the reader is referred to that
paper for details.  For the examples presented here, we used the
structure of the BW at the age of Tycho, as extracted from a
simulation where the SN Ia explosion model was a delayed detonation
with a kinetic energy of $1.16\times10^{51}$ erg \citep[model DDTc
from][]{bad05} - but such details do not affect the results for the
profile of thermal emission behind the BW.

In Figure~\ref{rimprof_therm} (top panel) we show two profiles that
span the range of observed shell thicknesses overlaid with the
profiles from these calculations.  The model profiles are not
limb-brightened on the scale presented and do an exceptionally poor
job of describing the observations.  Given that the geometric shell
models are such a good description of the data, only a modest fraction
of the rim emission could come from a component that is distributed
morphologically like shocked AM.  In lieu of a detailed morphological
model that explains the observed profile, as a first approximation, we
use a joint model consisting of the sum of the profiles of the thermal
emission and a purely geometric thin shell.  The bottom panel of
Figure~\ref{rimprof_therm} shows our joint model fits with the allowed
upper limit (90\% confidence) to the thermal emission from the shocked
AM shown as the lower dashed histograms.  For the somewhat thicker
profile (left side) as much as 33\% of the integrated rim emission
could come from the same thermal component.  This limit is reduced to
only 9\% for the thinner profile (right side).

\section{Tycho's Reverse Shock \label{rs}}

Although our focus has been on the location of the BW and CD and the
emission from the region between them, we wish to round out our
investigation of the fluid discontinuities in \ty\ by locating the
reverse shock (RS).  To do this, we take our cue from models of SN Ia
explosions.  SNe Ia are thermonuclear explosions involving C and O
white dwarfs \citep{hillnie}.  For many years, theoretical models of
SN Ia explosions relied on one-dimensional calculations, involving
stratification in the ejecta: Fe-peak elements are interior to Si, S
and other intermediate mass elements.  More recently,
three-dimensional models of SN Ia explosions have appeared, which
result in a very efficient mixing of all the elements throughout the
ejecta \citep{rei02,gam03,gar05}.  However, it is unclear whether
these models can explain the fundamental properties of the optical
spectra of SNe Ia \citep{bar03,bra04}, as well as reproduce the X-ray
spectra of SNRs.  Specifically, they predict similar spectral
properties for the Si and Fe emission in the shocked SN ejecta
\citep{bad05}.  This is in conflict with the results of \citet{hwa98}
who showed from \ty's integrated X-ray spectrum that a component of Fe
emission with a hotter temperature than that of the bulk Si emission
was required to reproduce the flux in the Fe K$\alpha$ line.
Additionally, observations show that this Fe K$\alpha$ emission peaks
interior to that of Fe L and Si \citep{hg97,dec01,hwa02}, suggesting
that there is a temperature gradient through the ejecta (cool near the
contact discontinuity, hottest near the reverse shock).  \citet{bad05}
also provided a theoretical interpretation for the hot Fe in the
interior, with incomplete collisionless electron heating at the
reverse shock and the subsequent evolution of temperature toward the
CD in SNe Ia.  Based on this evidence, we take the point of view that
the Fe K$\alpha$ emission comes from the innermost portion of the
shocked ejecta and its inner edge therefore denotes the location of
the reverse shock (RS).

Figure~\ref{fekimg} (bottom right) shows the continuum (4--6 keV band)
subtracted Fe K$\alpha$ line image (which contains $\sim$15000 counts
in total).  Over all but a portion of the eastern side, the image is
circularly symmetric and clearly limb-brightened. We extracted radial
profiles from within 12$\degr$-wide wedges as a function of azimuthal
angle around the entire SNR and fit each profile independently with
simple projected shell models.  These models are purely geometric and
assume a spherically-symmetric, uniform density shell.  The inner and
outer radii are the relevant fitting parameters.  Fits with 1, 2, and
4 shell components were tried; the location of the innermost radius,
i.e., the location of the RS, was insensitive to the number of
components.  The two-component shell model resulted in generally good
fits ($\chi^2_{reduced}\sim 1$) for azimuthal angles over
0$\degr$--48$\degr$ and 168$\degr$--360$\degr$ with modest
uncertainties in the RS radius. We find an average (error weighted)
radius for the RS of 183$\arcsec$. 

The ratio of the RS to BW radii over these azimuthal ranges is 72\%
and only increases to 73\% if we use the average BW radius from the
entire 360$\degr$.  In the following we will assume that the average
radius for the RS can be extrapolated to the full 360$\degr$.
Figure~\ref{radplot} shows the RS radius as a function of azimuthal
angle as the purple histogram, with 90\% error bars overplotted.  The
innermost contour on Figure~\ref{fekimg} (bottom right) indicates the
location of the RS.  It clearly traces the peak of the Fe K$\alpha$
line surface brightness around the remnant, which agrees with what is
expected for the projection of a thin shell.  In the east, although
the shell model describes the profiles accurately, the inner radius is
not well constrained by the fits and the errors on this quantity are
much larger than elsewhere (factors of 2 or more).  We have chosen to
exclude these regions from our further analysis.  The coincidence of
the bright Fe/Si knots and the seeming ``gap'' in the RS on the
eastern limb has not escaped our attention. We will address this
curious morphology in future work.

\section{Discussion}

Three key results on Tycho's SNR that address the spectral character
of the blast wave emission, its fine scale morphology, and the radial
positions of the blast wave, contact discontinuity, and reverse shock
emerge from the preceeding sections. Here we consider the implications
of these results for the nature of Tycho's SNR and show that all three
observables are consistent with a single specific intrepretation.

\subsection{Biases and Errors in the Locations of the Fluid Discontinuities}

Before proceeding to use the relative radial positions of the fluid
discontinuites to constrain the dynamical state of the remnant, we
must first be certain that the projected radii we measure are unbiased
estimates of the ``true'' radii.  Let us first consider the effect of
geometric projection.  If the blast wave and ejecta were perfectly
smooth and symmetric shells of emission, then the radii we observe
would be close to the true radii.  In practice, this is not the case;
the shells of emission are highly structured.  The true average radius
lies somewhere between the maximum and minimum radial extent of the
protrusions and indentations around the shell.  Projecting a highly
structured shell onto the plane of the sky tends to favor protruding
parts of the shell.  Therefore, the average radius we measure in
projection is an overestimate of the true average radius.

We have done simple simulations using our power-spectrum analysis (\S
\ref{powerspectrum}) to quantify the amount of bias from projection
for the BW and CD radii.  According to these simulations, our observed
values for both positions are over-estimated by $\sim$3\% and
$\sim$6\%, respectively.  Another estimate for the bias in the CD
radius comes from examining the hydrodynamical evolutionary models of
\citet{wc01}.  In these two-dimensional calculations, the
Rayleigh-Taylor instability at the CD allows fingers of ejecta to
protrude beyond the average CD radius.  \citet{wc01} find that the
maximum extent of these fingers can extend to 87\% of the BW radius,
while their one-dimensional simulations place the CD at 77\% of the BW
radius. The true average radius of the CD should lie between these
extremes. If the average CD radius we measure in projection is
dominated by the maximum extent of the fingers, then we would be
over-estimating the true CD radius by $\sim$5\%.

The RS radius we quote from the shell fits to the Fe K$\alpha$ image
is also likely to be over-estimated.  We know that there is Fe
K$\alpha$ emission at least down to the radius we quote; however,
there may be even more shocked ejecta further in that is simply too
faint to detect. Specifically our main concerns are the amount of
radial variation in the intrinsic Fe K$\alpha$ emissivity through the
ejecta shell (we assume uniform emissivity in our fits), and the
precise location of the reverse shock relative to the Fe K$\alpha$
emission.  To assess these effects we again turn to the hydrodynamical
simulations presented in \S \ref{morph}, only now we consider the Fe
ejecta emission. From the simulations we produced model Fe K$\alpha$
surface brightness profiles under a range of conditions.  The profiles
were generated on radial bins and with simulated errors to match the
data (e.g., lower panel of Figure~\ref{radprof}).  These were then fit
with the uniform shell model in the same way as the data profiles.
Comparing the fitted inner edge of the shell model with the precisely
known location of the RS from the simulation allowed us to estimate
the bias.  As expected the shell fits over-estimated the true radius
of the RS, but the bias factor was modest: it varied from 1\%--3\% for
plausible model conditions up to a factor of 8\% for a more extreme
case.  To be conservative we use a RS bias factor of 5\%.

We are now in the position to convert our measured projected average
radii for the BW, CD, and RS
(251$\arcsec$~:~241$\arcsec$~:~183$\arcsec$ or equivalently
1~:~0.96~:~0.73 when scaled to the BW) to relative values that more
accurately reflect the ``true'' radii of these surfaces.  As shown
above the relative bias between the CD and BW is $\sim$3\%, and
between the RS and BW the bias is $\sim$2\%.  Including these
correction factors, our best estimate for the true relative radii
become 1~:~0.93~:~0.71 (BW~:~CD~:~RS).  We use these values for the
remainder of the discussion. (Note that \citet{sew83}, using only the
\einstein\ broadband X-ray image, derived qualitative estimates for
the locations of the fluid discontinuities that are similar to the
values from our detailed analysis.)

One final point about the radii concerns their uncertainty.  The error
on the BW radius is negligible compared to the errors on the radii of
the CD and RS. An estimate for the error on the CD radius comes from
comparing the 2-$\sigma$ and 4-$\sigma$ average radii (see
\S\ref{bwandcd}) with the nominal 3-$\sigma$ value.  This prescription
gives an error on the relative CD radius of $\pm$2\%. Our shell fits
resulted in uncertainties on the RS radii for each of 20 separate
12$^\circ$ wide azimuthal sectors that varied from 4$^{\prime\prime}$
to 17$^{\prime\prime}$. If each sector is assumed to provide an
independent estimate for the RS radius, then the formal error on the
average RS radius would be $\sim$1$\farcs$4, which is rather
optimistic.  To be conservative we take the average uncertainty from
the separate fits, 7$\farcs$6.  This corresponds to an error on the
relative RS radius of $\pm$3\%.

\subsection{Adiabatic Hydrodynamical Models}

There are only a handful of references in the literature that present
numerical values for all three of the radial locations we have
measured in \ty. \citet{ham86} found relative radii of 1~:~0.86~:~0.83
at the dynamical age of Tycho for uniform density ejecta in an
undecelerated blast wave, while \citet{wc01} found 1~:~0.77~:~0.66 for
ejecta with an exponential density profile.  Neither of these
simulations comes close to matching our observed CD:BW ratio of 0.93.
Since the ratio of radii depends on the density profiles in the ejecta
and ambient medium and additionally varies with time for a given set
of profiles, we perform a more detailed comparison between our
observed values and the predicted radii for the BW, CD and RS from the
hydrodynamical simulations of \citet{bad03} and \citet{bad05}.  These
models utilize realistic density profiles for the ejecta obtained from
SN Ia explosion models.  The ambient medium is assumed to be uniform.

The three curves in Figure~\ref{ratddtc} show how the ratio of radii
vary with time for one specific SN Ia explosion model \citep[case DDTc
from][]{bad05}. The horizontal axis in this figure is the normalized
time, $t^\prime$. To obtain the actual physical age of the remnant one
multiplies $t^\prime$ by the factor

\begin{equation}
T^\prime = { M_{ej}^{5/6}\over [(4\pi/3)\rho_{AM}]^{1/3}(2 E_K)^{1/2}}
\end{equation}

\noindent
\citep[eqn.~5 of][]{bad03}.  The ejected mass $M_{ej}$ and initial
kinetic energy $E_K$ are fixed by the SN Ia explosion model and the
age of Tycho is well-known (431 yr).  So constraints on $t^\prime$
correspond to constraints on the allowed ambient density $\rho_{AM}$.
Another independent constraint on the dynamics comes from the measured
angular radius of the BW, which we use below to constrain the distance
to Tycho.

From Figure~\ref{ratddtc} we see that over no part of the evolution
does the predicted CD:BW ratio match the observations (i.e., the green
curve does not intersect the hatched region). The two other predicted
ratios for this particular model do match the observations and set
constraints on the allowed range of normalized time coordinate.
However, these matches occur for {\em different} values along the
normalized time coordinate.  Thus the RS:BW ratio requires one range
of values for the ambient density, while the RS:CD ratio requires a
different, non-overlapping range of values.  Clearly this model is
unable to properly explain the locations of the BW, CD, and RS in a
physically consistent way.

We repeated this study for the entire grid of one-dimensional
hydrodynamical calculations based on the SN Ia explosion models from
\citet{bad03}.  These models cover the full range of SN Ia explosion
scenarios in the literature including detonations, deflagrations,
delayed detonations, pulsating delayed detonations, and
sub-Chandrasekhar models.  In addition we include the well-known W7
model \citep{nom84} as well as a simple analytical exponential profile
\citep[e.g.,][]{dc98}.  We use the constraints on normalized time (as
in Figure~\ref{ratddtc}) and the angular size of the remnant to
determine the allowed range of ambient density and distance.  Results
are shown in Figure~\ref{distden}.  For each of these models, curves
from the RS:BW ratio (solid curves on left) and the RS:CD ratio
(dashed curves on right) do not overlap (same problem illustrated in
Figure~\ref{ratddtc}).  The constraints provided by these ratios are
physically inconsistent. These models all suffer from the same
problem: {\em they predict that the CD is much further from the BW
than we observe}.

The fact that the CD is so close to the BW while the RS remains deep
in the ejecta is very puzzling.  According to \citet{wc01},
Rayleigh-Taylor instabilities can produce fingers of ejecta that
protrude out toward the BW with the {\em maximum} extent of these
fingers approaching fractional radii of $\sim$0.87.  This is nowhere
near the value of 0.93 that we find for the {\em average} location of
the CD.

\subsection{Ejecta Clumps}

\citet{wc01} proposed that a set of high-density, high-velocity ejecta
clumps in Tycho can cause the ejecta to appear to be quite close to
the BW. In order to survive as distinct features near the BW at
present, these clumps must have formed with a high density contrast
($\gtrsim$ 100) and velocities of $\sim$7000 km s$^{-1}$ during the
explosion. These authors suggest the Ni bubble effect as a plausible
mechanism for forming such clumps, but detailed calculations of the
process have yet to be done.  If a limited number of such clumps (we
find of order ten or so that protrude to the BW in projection, see
Figure~\ref{radplot}) have skewed the location of the CD to larger
radii, then a more appropriate estimate for the bulk of the ejecta
would exclude the clumps.  We therefore also calculated the average CD
radius for two angular regions with no obvious ejecta protrusions: for
203$\degr$--270$\degr$ we find the ratio of CD to BW radii to be 0.92,
and for 292$\degr$--329$\degr$ we find a ratio of 0.91 (taking account
of projection).  These values are still too large to match the
evolutionary models.  Additionally, although detailed calculations of
the X-ray emission have not been published, the relevant properties
(temperature, ionization state, etc.) of the clumps at the edge of the
SNR should differ from the bulk of the ejecta. However at least from
Figure~\ref{color}, we have found no obvious morphological or
broadband spectral differences between the protruding ejecta and that
in the interior. An upcoming article \citep{hw05} addresses the
observed properties of the ejecta in Tycho.  Finally this scenario
explains neither the featureless nature of the blast wave spectra nor
the thin rim morphology.

\subsection{Cosmic Ray Acceleration}

Our preferred explanation for the location of the BW, CD, and RS
invokes the presence of efficient cosmic ray acceleration in \ty.
Cosmic ray acceleration at the BW would increase the compression
factor there (above the usual value of 4) with a concomitant shrinking
of the gap between the BW and CD \citep{dec00,be01}.  In particular,
\citet{be01} state that if the compression factor is large enough,
fingers of ejecta resulting from Rayleigh-Taylor instabilities at the
CD can reach into the shocked interstellar medium and perturb the BW,
as we observe in \ty.  In the following we divide the discussion into
evidence for cosmic ray acceleration of electrons and ions at the BW,
and evidence for cosmic ray acceleration at the RS.

\subsubsection{Evidence for Acceleration of Electrons \label{elacc}}

If the BW in SNRs can accelerate cosmic ray {\em electrons} to high
energies, the electrons will emit X-ray synchrotron radiation.  We
find two pieces of evidence for this in \ty.

\paragraph{Spectra}

We have shown the spectra of three regions around the rim of \ty\ that
are {\em extremely} featureless (Figure~\ref{spectra}). When fitted by
an absorbed power-law model, we find photon indices ($\alpha_p
\sim$2.7) that agree with the \ginga\ best fit value.  The index is
similar to that found in other SNRs with non-thermal X-rays
\citep[e.g., $\alpha_p \sim$2.95 for SN 1006,][]{koy95}.

On the other hand, a thermal interpretation for the rim emission
cannot be rejected based purely on spectral analysis. \citet{hwa02}
and \citet{bam} claim that both thermal and non-thermal models provide
statistically indistinguishable quality of fits to their data. To
describe the rim spectra, line emission in the thermal models must be
suppressed. There are two ways to accomplish this: the emitting plasma
must be severely underionized or it must have very low abundances.
\citet{hwa02} examined the former case and obtained a limit on the
ionization timescale (the quantity that characterizes the ionization
state of the post-shock plasma) of $n_et\sim10^8$ cm$^{-3}$ s.  This
value is remarkably low and, if true, would be unprecedented.
However, we can show that this scenario is internally inconsistent.
Knowing the angular expansion rate of the remnant
\citep[$\sim$0.3$\arcsec$ yr$^{-1}$;][]{hugh}, one can determine the
time it took the plasma to flow across a given post-shock distance.
The interior rim regions fitted by \citet{hwa02} lie $\sim$5$\arcsec$
in from the BW, which extrapolates to a flow time of $\sim$60 yrs
(with respect to the shock).  The electron density therefore needs to
be of order $\sim$0.05 cm$^{-3}$ to produce a very low ionization
timescale. One can also obtain an estimate of the density from the
fitted normalization of the spectrum and an estimate of the emitting
volume.  Using our rim spectra we obtain densities in this manner of
$\sim$$10\, (D/2.3\, {\rm kpc})^{-1/2}$ cm$^{-3}$.  The two density
estimates differ by more than two orders of magnitude, which is far
beyond any possible hope of redemption or accommodation by tweaking
volume estimates, flow times, or other quantities.

Given the intensity of the emission, a more reasonable value for the
ionization timescale would be $n_et\sim2\times10^{10}$ cm$^{-3}$
s. With this value fixed, we estimated the factor by which the metal
abundances have to be reduced to obtain sufficient suppression of the
line emission.  We find that the abundances must be $<$3\% of the
solar values in order to describe the data. Although it is possible
that certain species might be depleted onto dust grains (i.e., the
refractory elements like Si and Fe), neon, as it is inert, should be
present in the gas phase and should produce easily visible line
emission around 1 keV.  All things considered a thermal interpretation
for the featureless spectra at the BW in \ty\ must be considered an
extremely unlikely possibility.

The integrated \ginga\ spectrum of \ty\ \citep{fink94} shows a hard
power-law component that entirely dominates the emission above
$\sim$15 keV.  Our integrated \chandra\ spectrum over the 4--9 keV
band (excluding the Fe K$\alpha$ line) is consistent with the
extrapolation of the best-fit \ginga\ spectral model both in terms of
shape and normalization (the best fit normalization from \chandra\ is
only 4\% higher than the quoted \ginga\ value).  In this \chandra\
band, the extrapolated powerlaw component accounts for about 60\% of
the flux with an additional 40\% coming from a thermal component
(again assuming the best-fit \ginga\ spectral model). The presence of
a hard featureless component in \ty's X-ray spectrum is thus
reasonably secure.  As we have argued above with the rim spectra, this
hard featureless component of emission likely arises from the BW in
\ty.

\paragraph{Morphology \label{magfield}}
Portions of the rim that show featureless spectra also show thin
filaments of limb-brightened emission in the 4--6 keV band in our
\chandra\ data. These features are nearly ubiquitous around the entire
rim.  We fit models of the predicted morphology assuming a
shock-heated thermal origin for the emission profiles (see \S
\ref{morph} and {Figure~\ref{rimprof_therm}).  The thermal profiles
cannot reproduce the sharp brightness contrast between the rim
filaments and the region immediately behind the rim, while a simple
projected thin shell model does provide a good fit (see \S \ref{rim}
and Figure~\ref{rimprof}).

The acceleration of cosmic ray electrons gives a natural explanation
for a thin emitting region: electrons will lose energy as they emit
synchrotron radiation, eventually ceasing to emit X-rays.  In
addition, the acceleration mechanism, believed to be diffusive shock
acceleration (DSA) \citep[see e.g.,][and references therein]{rey98},
constantly scatters cosmic rays across the shock front.  A particle
scattered or advected a certain distance from the shock must diffuse
back across it in order to be further accelerated.  If this distance
is longer than the diffusion length, the particle will lose its energy
via synchrotron radiation before reaching the shock.  These two
effects, synchrotron losses and diffusion, combine to explain why
X-ray synchrotron emission is seen in only a thin region behind the
shock \citep[for an alternate view see][]{pohl}.

Since that is the case, the widths of the rims in \ty\ can be used to
estimate the post-shock magnetic field, $B$.  To do this, we follow
the method of \citet{bv04}, who solve a one-dimensional transport
equation including diffusion, bulk flow, and loss terms to get a
relation for the magnetic field:

\begin{equation}
B = 141 \mu {\textrm G}
\left[\left(\frac{\Delta\theta}{2\arcsec}\right)
\left(\frac{D}{2.3\,\textrm{kpc}}\right)\right]^{-2/3}
(\sqrt{1+\delta^2}-\delta)^{-2/3},
\end{equation}

\noindent
where

\begin{equation}
\delta^{2}=0.026
\left(\frac{\sigma}{7}\right)^{-2}
\left(\frac{h\nu_{max}}{5\,\textrm{keV}}\right)^{-1}
\left[\left(\frac{p}{0.124\,\textrm{\% yr}^{-1}}\right)
\left(\frac{D}{2.3\,\textrm{kpc}}\right)
\left(\frac{\theta_{BW}}{243\arcsec}\right)\right]^2 .
\end{equation}

\noindent
We use the average X-ray expansion rate, $p$, of 0.124\% yr$^{-1}$
from \citet{hugh}.  Although this differs significantly from the
0.113\% yr$^{-1}$ radio measurement of \citet{reynoso}, more recent
radio expansion measurements \citep{mof04} indicate a rate of 0.123\%
yr$^{-1}$, in better agreement with the X-ray result.  The expansion
rate we use corresponds to a shock speed of $3400(D/2.3\, {\rm kpc})$
km s$^{-1}$. For the angular radius of the blast wave, $\theta_{BW}$,
we use the projection-corrected value of $243\arcsec$.  We use a
maximum photon frequency, $\nu_{max}$, corresponding to 5 keV, since
this is the middle of the 4--6 keV continuum emission used to create
the rim profiles.  The shock compression factor, $\sigma$, is taken to
be 7, the value for a relativistic gas (the dependence on $\sigma$ is
weak, in any case). We use $\Delta\theta$ for the shell widths for
both cases (i.e., uniform and exponential emissivity profiles).  For
the uniform profile widths, we find $B=78-158(D/2.3\,{\rm
kpc})^{-0.55}$ $\mu$G.  The widths from the exponential profile give
$B=123-436(D/2.3\,{\rm kpc})^{-0.55}$ $\mu$G.  Note that the distance
scaling is approximate; the actual dependence on distance is more
complicated, but these scalings are accurate for the usual range of
1.5--3 kpc.  \citet{vbk05} calculated \ty's magnetic field in this
manner.  They used widths from an exponential profile and slightly
different values of $p$, $\nu_{max}$, and $\sigma$.  They obtained
values of 150--373 $\mu$G, similar to what we find.

In the literature, there are other ways of determining the magnetic
field \citep[e.g.,][]{vl03,bal} that can be obtained from limit cases
of the method presented in \citet{bv04}.  If the acceleration is
efficient, then the synchrotron loss time is much longer than the time
to accelerate particles to energy $E=h\nu_{max}$.  In other words,
particles have time to diffuse across the shock before losing energy
via synchrotron radiation.  In this limit we recover the method of
\citet{vl03}:

\begin{equation}
B = 65.4 \mu {\textrm G} \left[
\left(\frac{\sigma}{7}\right)
\left(\frac{\Delta\theta}{2\arcsec}\right) \right]^{-2/3}
\left(\frac{h\nu_{max}}{5\textrm{keV}}\right)^{-1/3}
\left[\left(\frac{p}{0.124\textrm{\% yr}^{-1}}\right)
\left(\frac{\theta_{BW}}{243\arcsec}\right)\right]^{2/3} .
\end{equation}

\noindent
We find $B=33-66$ $\mu$G for the uniform profile widths and $B=51-182$
$\mu$G for the exponential profile widths.  This method is
distance-independent.  On the other hand, the acceleration time may be
longer than the synchrotron loss time, and a particle will have no
time to diffuse back across the shock before losing its energy via
synchrotron radiation.  In this limit, $\delta^2 = 0$ in eqn.~3, and
we recover the method of \citet{bal}.  We find the uniform profile
widths yield $B=70-142(D/2.3\, {\rm kpc})^{-2/3}$ $\mu$G, while the
exponential profile widths yield $B=110-392(D/2.3\, {\rm kpc})^{-2/3}$
$\mu$G.  This method does not depend on $\sigma$.  \citet{bal} carries
out this calculation for \ty\ and finds $B=250$ $\mu$G, using a single
width from \citet{hwa02}.

The magnetic field estimates from the uniform profile widths are
consistently 1.5--2.7 times lower than estimates from the exponential
profile widths.  A more accurate determination of the magnetic field
depends on understanding the true emissivity profile of the rim, yet
we have shown that both the uniform and exponential profile models
provide equally good fits to the rim emission (see \S \ref{rim}).
Additionally, the three methods for finding $B$ rely on different
assumptions about the synchrotron loss time compared to the time for
particles to be accelerated to relativistic energies.  It is beyond
the scope of this paper to investigate which is most appropriate, yet
it is clear further work must be done to provide accurate estimates of
the magnetic field strength in SNRs.  A promising avenue would be to
compare the radio and X-ray rim morphologies \citep[e.g.,][]{elcc}.

One possible argument against efficient cosmic ray acceleration at the
forward shock of \ty\ is the presence of Balmer-line filaments
distributed non-uniformly around the rim in the north and east.  These
filaments come from non-radiative shocks propagating into a partially
neutral AM \citep{chev}.  According to \citet{draine}, this neutral
component reduces the cosmic ray acceleration efficiency.  Although
most of the Balmer-line filaments do {\em not} appear to coincide with
the 4--6 keV rim filaments, their mere existence suggests that the
efficiency of cosmic ray acceleration varies around the rim of \ty.
We leave a detailed investigation of this to a future study.

\subsubsection{Evidence for Acceleration of Ions}

Our estimates for the post-shock magnetic field strength (see \S
\ref{magfield}) in \ty\ range from 33 to 436 $\mu$G.  This range comes
nearly equally from three sources: (1) the different observed widths
around the rim, (2) the different methods for determining these widths
(uniform vs.~exponential emissivity profiles), and (3) the different
assumptions made about the diffusion and loss times.  Making the most
conservative assumptions for items (2) and (3), but considering the
thinnest portion of the rim around \ty\ (item 1), we find a
``minimum'' pre-shock magnetic field strength of $\sim$10 $\mu$G
(assuming a shock compression factor of 7).  This is a factor of 2
times the canonical 5 $\mu$G value for the ISM \citep[e.g.,][]{ferr},
which is evidence that amplification of the magnetic field is
occurring at the forward shock of Tycho.  \citet{vbk05} state that
such amplification can happen if a significant portion of the cosmic
rays are ions.

In this article we have presented strong evidence that the CD and BW
in \ty\ are too close together to be described by standard adiabatic
hydrodynamical models, regardless of the density or velocity structure
of the SN ejecta or the evolutionary state.  The closeness of the CD
and BW in \ty\ points to a higher compression factor for the forward
shock \citep{el05}, which is then evidence for efficient cosmic ray
acceleration.  Since the ions dominate the mass of the fluid, these
statements actually say more about the nature of the ions than the
electrons.  In particular, they point to the presence of relativistic
ions, as we show next.

The pressure that relativistic electrons {\em alone} exert is
insufficient to significantly alter the dynamics at the forward shock.
The total available pressure at the shock can be approximated by its
ram pressure, $\rho_{AM}V_{shock}^2=1.9\times10^{-7}$erg cm$^{-3}$
(for $\rho_{AM}=1.67\times10^{-24}$ g cm$^{-3}$ and $V_{shock}=3400$
km s$^{-1}$).  To have a significant affect on the dynamics, the
pressure in relativistic particles should be of order $\sim$50\% of
this ram pressure \citep[e.g., Figure 1 of][]{dec00}.  We calculate
the pressure in relativistic electrons in \ty\ following the method of
\citet[][p.~292]{long}.  We use a radio spectral index of 0.61 and a
flux density at 1 GHz of 56 Jy \citep{green}, our magnetic field
estimates (from \S \ref{magfield}), and a distance of 2.3 kpc.
Assuming the electrons fill the volume between the BW and CD, we find
that the electrons provide $<$0.23\% of the ram pressure, for any of
our allowed magnetic fields.  Some of the radio emission comes from
the thin rims (although the bulk is from the larger volume just used),
which are also where we estimated the magnetic fields.  If we assume
the volume of a shell of width 4$\arcsec$ (corresponding to the width
of the thin synchrotron filaments at the rim), the relative electron
pressure increases to only $<$0.9\%.  This is clearly not enough to
affect the dynamics of the forward shock; something else must be
providing the necessary pressure.

To include cosmic ray protons in our calculations, we assumed that the
energy in particles is the sum of the energies in protons and
electrons, and further, that the energy in protons is some factor
times the energy in electrons \citep[][p.~292]{long}.  For the
pressure in relativistic particles to be 50\% of the ram pressure, the
energy density in protons must be $\ge 50$ times the energy density in
electrons.  This is the most conservative estimate, using our lowest
magnetic field value (33 $\mu$G) and the volume of a shell of width
4$\arcsec$.  Taken together with the evidence for high energy
relativistic electrons at the BW in \ty\, there is now a compelling
case for cosmic ray acceleration of {\em both} ions and electrons at
the forward shock in Tycho.  We further note the broad consistency
between our estimate of the ratio of proton to electron energy
densities in \ty\ and the well-established value for Galactic cosmic
rays of $\sim$100 \citep[p.~307]{long}.

\subsubsection{Evidence for Acceleration at the Reverse Shock}

Some authors have proposed that the RS can also accelerate cosmic
rays.  If this occurs, then the gap between the RS and CD will also
shrink \citep{dec00}.  However, our results argue against a
significant increase in the compression factor at the RS.  The
observed gap between the RS and CD is rather large.  \citet{el05} have
investigated particle acceleration at the RS and find that increased
compression factors there result in much lower post-shock temperatures
than in the case with no acceleration. They state that this prediction
is a problem for SNRs with strong Fe K$\alpha$ lines, as we see in
\ty.  These authors also point out that particle acceleration at the
reverse shock in remnants of Type Ia SNe is expected to be negligible
unless the original magnetic field of the progenitor white dwarf was
unusually large or amplification of the field occurred subsequently.
Additionally, \citet{dec00} fit models to the spectrum of Kepler's SNR
and found a good fit for cosmic ray acceleration at the BW, but no
acceleration at the RS.  This is consistent with our findings for \ty.

\section{Summary}

We have used the relative locations of the forward shock, contact
discontinuity, and reverse shock to argue for efficient cosmic ray
acceleration at \ty's forward shock.  The ratio of the CD to BW radii
(0.93) is inconsistent with adiabatic hydrodynamical models, even when
Rayleigh-Taylor instabilities are taken into account.  The closeness
of the CD and BW are consistent with a scenario in which the forward
shock is efficiently accelerating cosmic ray ions.  This causes the
shock compression factor to increase, and the gap between the CD and
BW to shrink.  Evidence for the acceleration of cosmic ray electrons
to TeV energies comes from the interpretation of the spectrally
featureless rim as nonthermal X-ray synchrotron emission.  A modest
fraction ($\sim$40\%) of the 4--6 keV continuum emission observed by
\chandra\ from the entire remnant is thermal and it is possible that
most of this is associated with the prominent thermally--emitting
ejecta.  At the rim, the fraction of 4--6 keV emission that can be
thermal is much smaller, $\sim$10\%.  In fact the non-thermal
interpretation by itself can explain the spectrum and morphology of
the rim and a 2 keV thermal model is not required.  

The question then arises: since the ambient density around \ty\ cannot
be exceptionally low, where is the thermal emission from the blast
wave?  Based on our work \citep[and also noted by][]{hwa02} the
electron temperature of the blast wave must be less than 2 keV.  In
the literature there is one measurement of the proton and electron
temperatures at the rim of Tycho's SNR from the optical Balmer line
emission of knot ``g'' at the eastern limb \citep{ghav}.  These
authors find a post-shock proton temperature of $\sim$10$^8$ K and an
electron temperature of $<$10$^7$ K ($<$0.9 keV) (although their
models assume negligible cosmic ray acceleration).  If the average
electron temperature around the rim of \ty\ is comparable to or
somewhat less than this value, its X-ray emission will be absorbed by
the ISM and therefore largely invisible to us.  Thus the rim of \ty's
SNR may be similar to other remnants, such as G347.3$-$0.5, in which
the X-rays appear to be purely nonthermal with no detectable thermal
emission present \citep{sla99}. In cases like these, where information
from the thermal plasma is lacking, other tracers of the dynamical
state of the remnant, such as expansion rates, locations of the fluid
discontinuities, and proton/electron temperatures measured from
Balmer-line filaments, become that much more precious.

For years, astronomers have been estimating explosion energies, ages,
and ambient densities for young SNRs based on the observed properties
of the forward shock under the assumption of adiabatic dynamics
\citep[e.g., Sedov-Taylor,][]{tm99}.  As we have shown here, the
dynamics of the forward shock in \ty\ are dominated by efficient
cosmic ray acceleration.  If \ty\ is a typical case, then all these
previous estimates are based on incorrect assumptions.  Future work
needs to concentrate on improving model assumptions, determining how
common the cosmic-ray-acceleration-dominated phase is and constraining
its duration in order to arrive at a better understanding of the
relationship between SNe and SNRs, SNR evolution, and the influence
they have on galactic environments.

\acknowledgments

JPH acknowledges support from the International Space Science
Institute in Bern, Switzerland, which facilitated fruitful discussions
with Anne Decourchelle, Don Ellison, and other colleagues on topics
related to the work presented here.  We thank David Burrows and John
Nousek for assistance with the original proposal, as well as Ken
Nomoto for the use of his W7 model and Jacco Vink for a helpful
discussion about magnetic field estimates.  CFM acknowledges support
from NSF grant AST00-98365 and from the Miller Foundation for Basic
Research. Support for this work was provided by the National
Aeronautics and Space Administration through \chandra\ Award Number
GO3-4066X issued to Rutgers by the \cxo\ Center, which is operated by
the Smithsonian Astrophysical Observatory for and on behalf of the
National Aeronautics Space Administration under contract NAS8-03060.

\clearpage

\begin{deluxetable}{ccc}
\tablewidth{0pt}
\tablecaption{Energy Bands used for PCA}
\tablecolumns{2}
\tablehead{\colhead{Number} & \colhead{Energy Range (keV)} &
\colhead{Feature}}
\startdata
 1 & 0.50--0.70 & O                    \\
 2 & 0.70--0.95 & Fe L                 \\
 3 & 0.95--1.26 & Fe L/Ne              \\
 4 & 1.26--1.43 & Mg/Fe L              \\
 5 & 1.43--1.63 & continuum            \\
 6 & 1.63--2.26 & Si K                 \\
 7 & 2.26--2.96 & S K                  \\
 8 & 2.96--3.55 & Ar K                 \\
 9 & 3.55--4.10 & Ca K                 \\
10 & 4.10--6.10 & continuum            \\
11 & 6.10--6.80 & Fe K                 \\
12 & 6.80--9.00 & continuum/background \\
\enddata
\tablecomments{Features listed are general, and only possible (not
necessarily detected) for that energy range.}
\label{bins}
\end{deluxetable}

\begin{figure}
\epsscale{.45}
\plotone{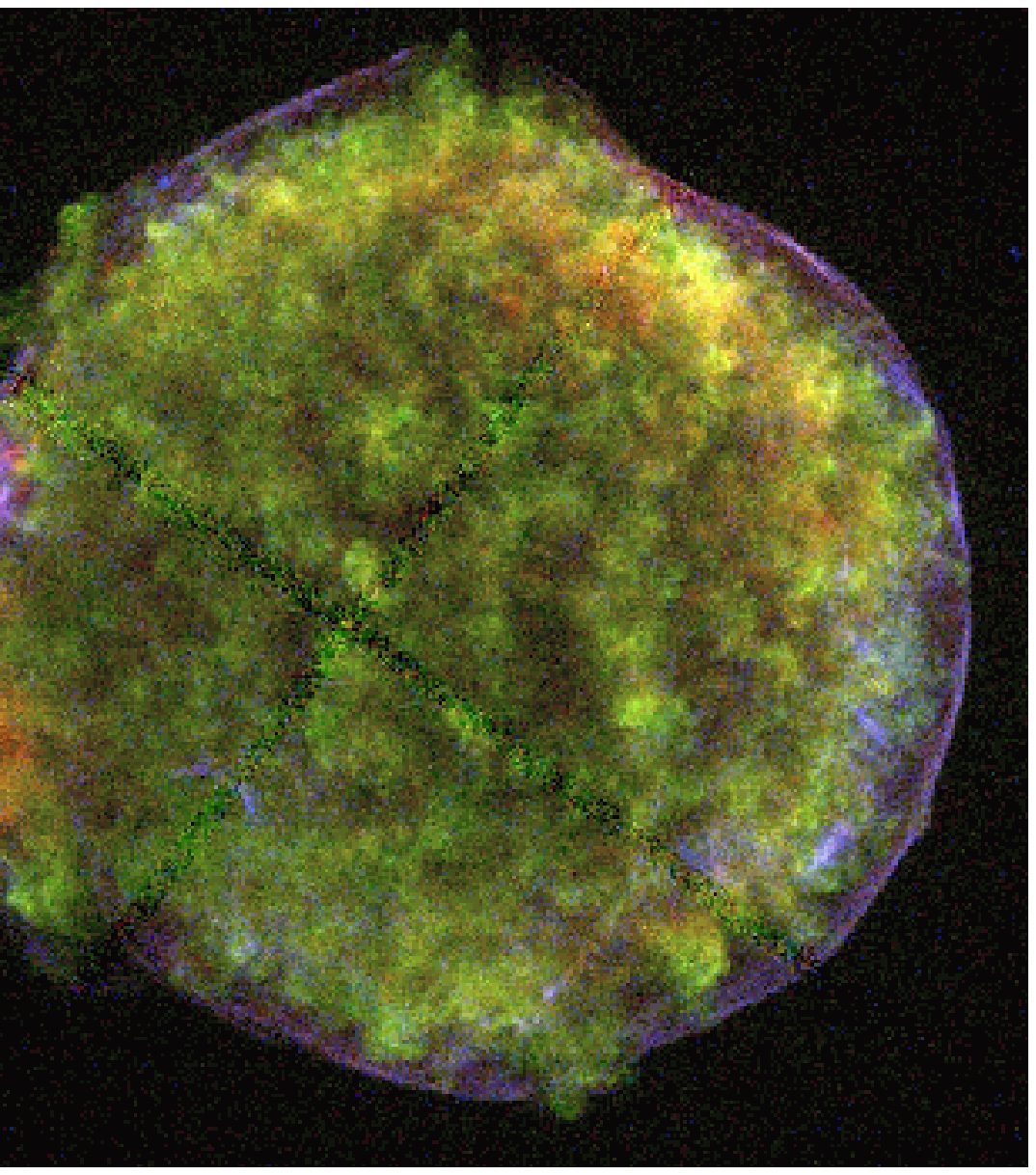}
\plotone{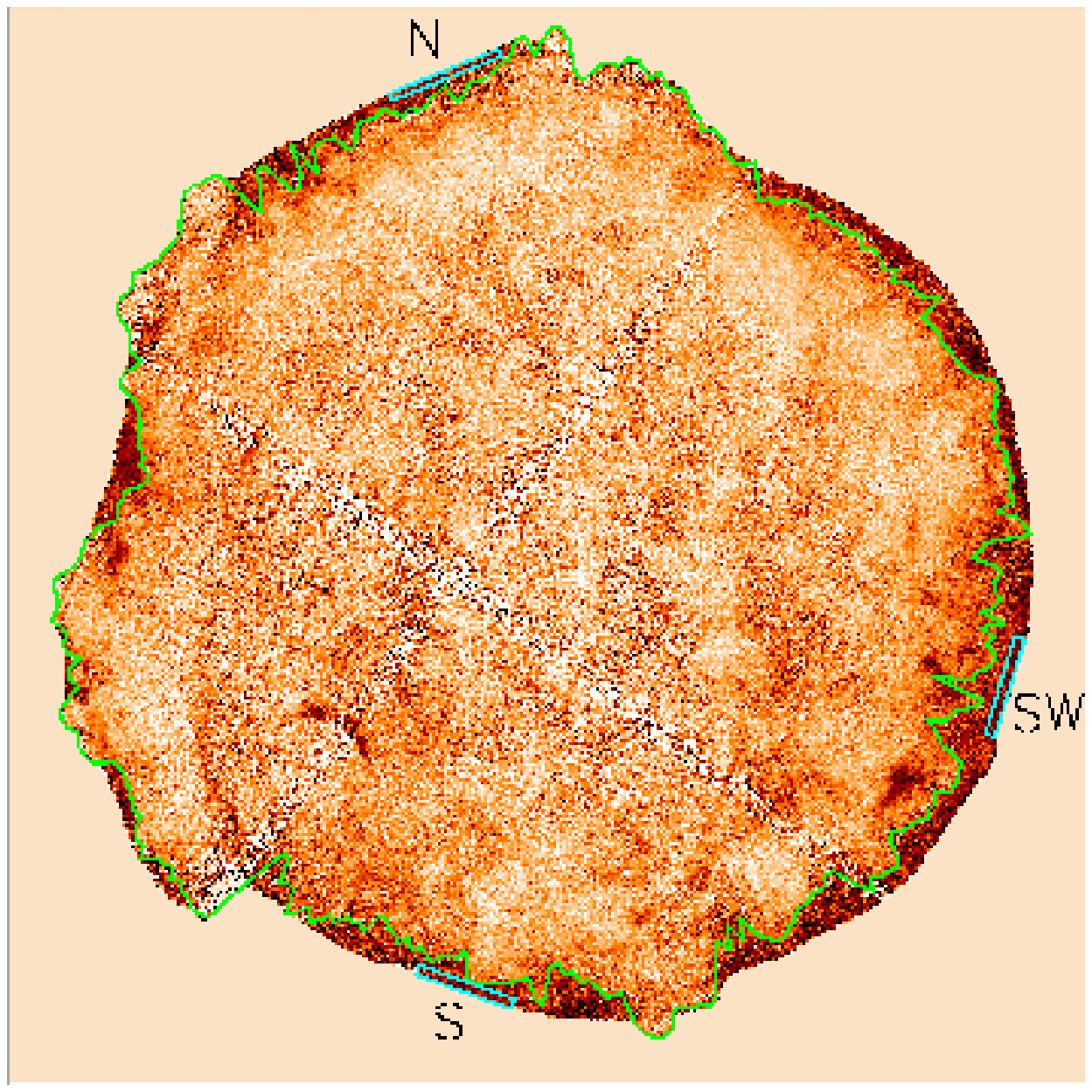}
\plotone{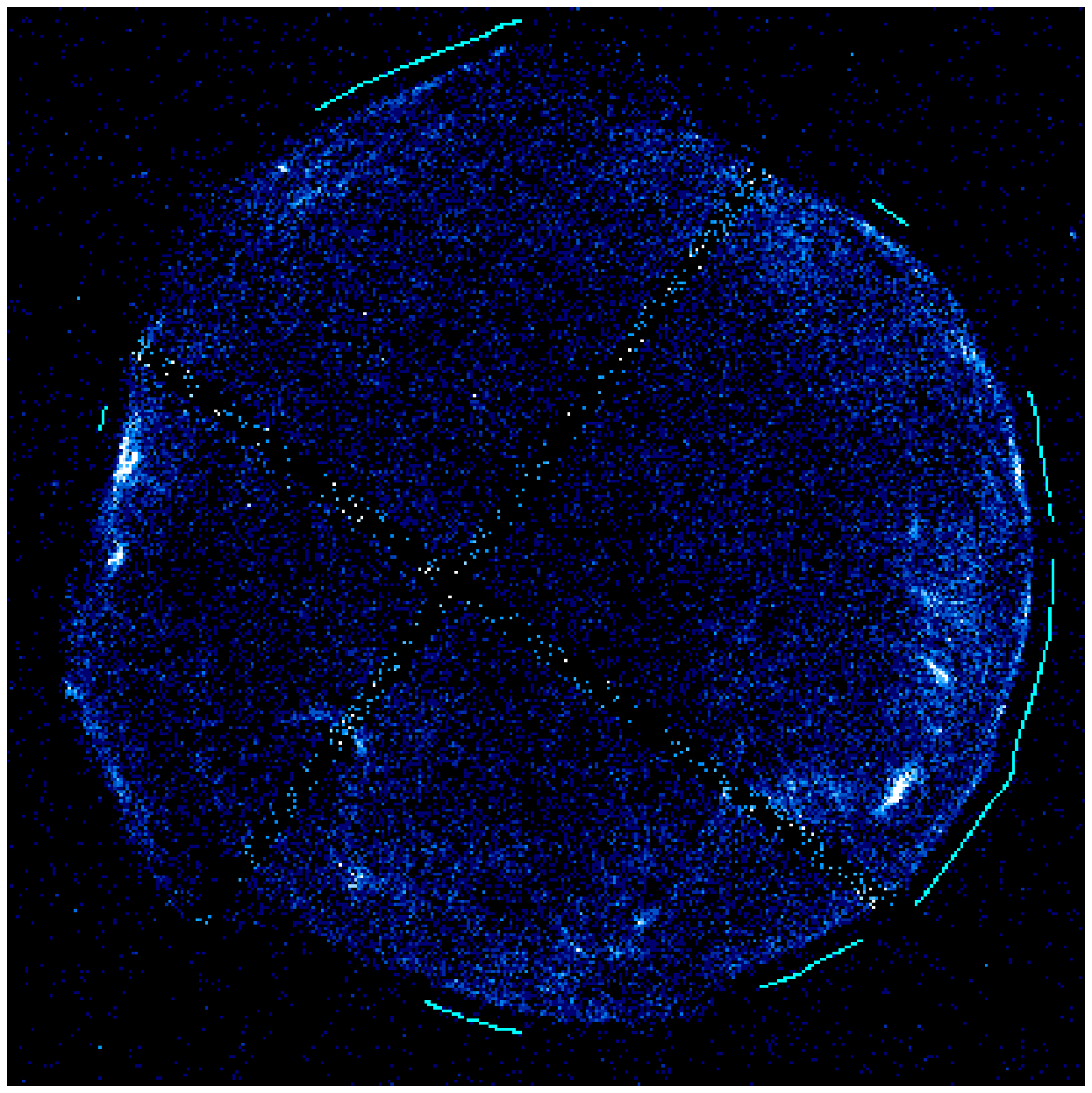}
\plotone{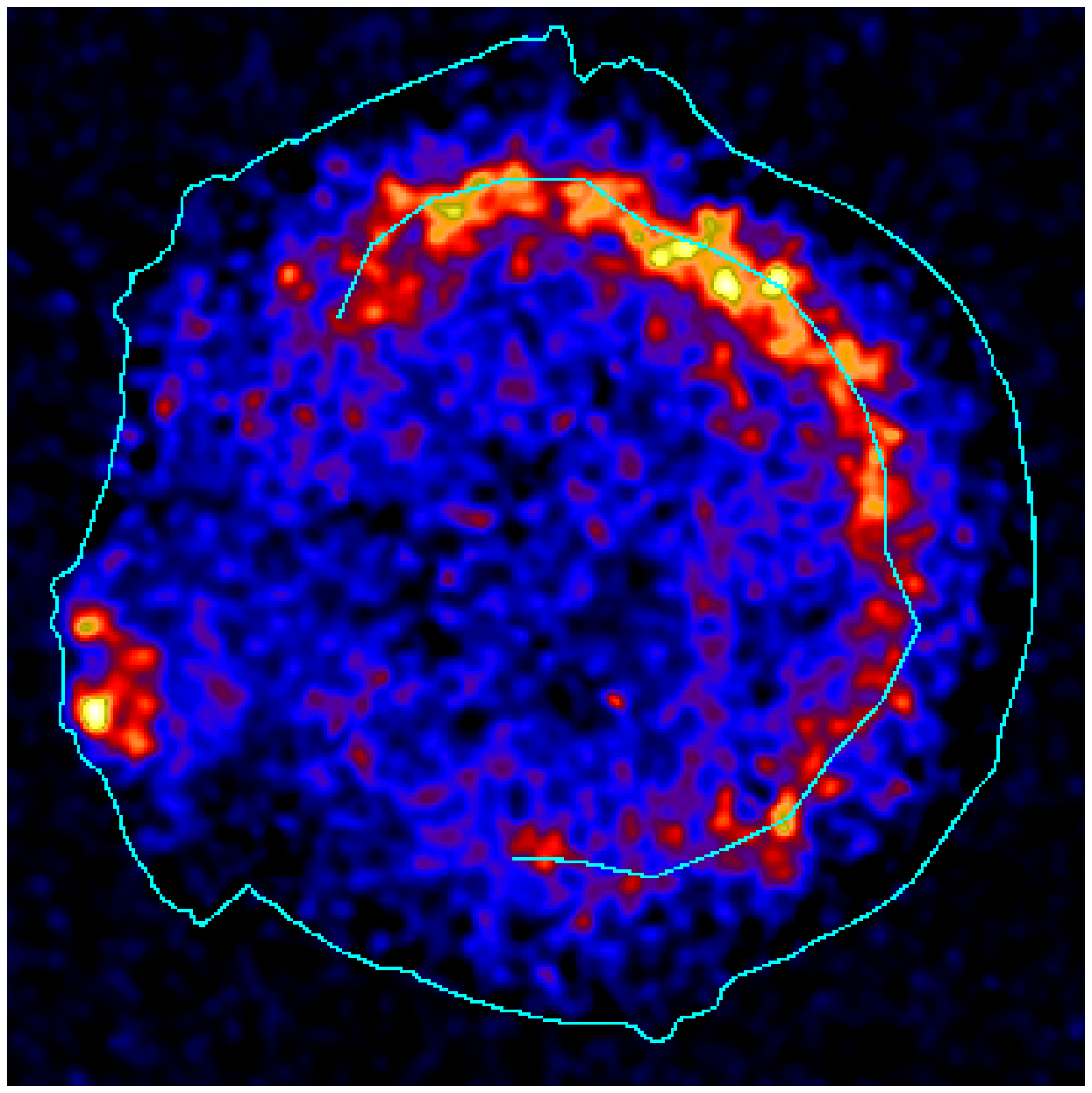}
\caption{{\em Top left}: Three-color composite \chandra\ image of
Tycho's SNR.  The red, green, and blue images correspond to photon
energies in the 0.95--1.26 keV, 1.63--2.26 keV, and 4.1--6.1 keV
bands, respectively.  {\em Top right}: An image of the first principal
component (PC1) that separates line-rich emission (light regions) from
featureless emission (dark regions). The green contour indicates the
location of the contact discontinuity. Three spectral extraction
regions are indicated.  {\em Bottom left}: Continuum (4--6 keV band)
image with regions used to determine width of rim filaments indicated.
{\em Bottom right}: Fe K$\alpha$ line image with continuum (4--6 keV
band) subtracted. The inner contour notes the location of the reverse
shock and the outer contour the location of the blast wave.  The field
of view of each panel is $9.5\arcmin \times 9.5\arcmin$.  North is up
and east is to the left.
\label{color}
\label{pc1}}
\label{contimg}
\label{fekimg}
\end{figure}

\clearpage
\begin{figure}
\epsscale{0.8}
\plotone{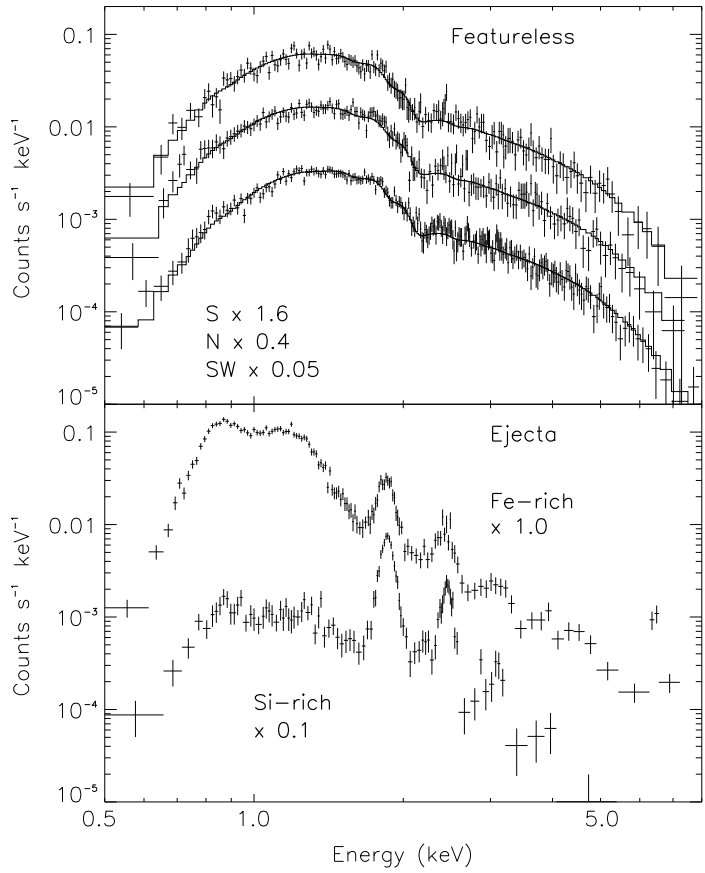}
\caption{{\em Top:} Spectra and best fit models for featureless
emission.  {\em Bottom:} Spectra for ejecta emission.  The Fe-rich
spectrum is from the reddish region in the southeast, and the Si-rich
region is from the yellowish region in the southeast.
\label{spectra}}
\end{figure}

\clearpage
\begin{figure}
\epsscale{1.0}
\plotone{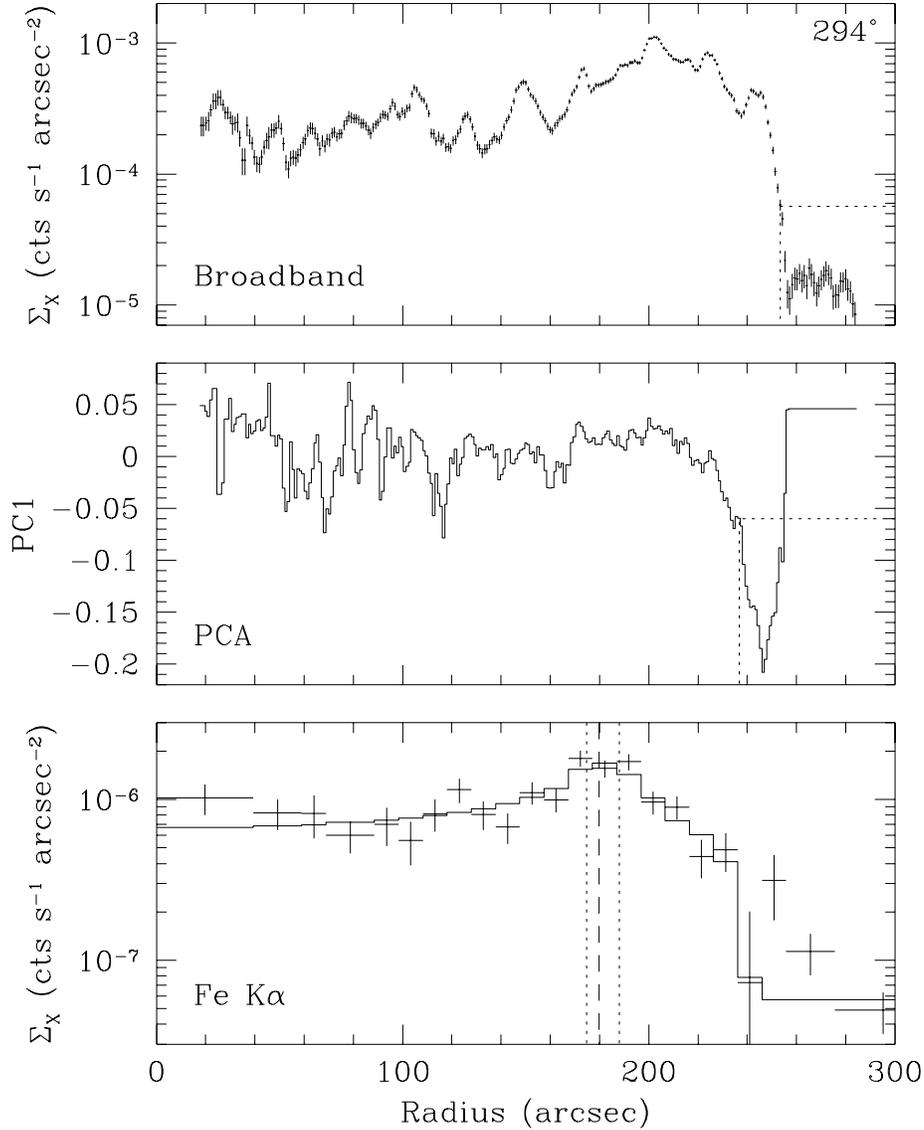}
\caption{Radial surface brightness profiles for a wedge centered at
294$\degr$, measured counterclockwise from north. {\em Top}: Broadband
emission.  The dashed lines indicate the contour level chosen to mark
the blast wave (5.7$\times10^{-5}$ cts s$^{-1}$ arcsec$^{-2}$) and the
corresponding radius.  {\em Middle}: PC1 values.  The dashed lines
indicate the threshold chosen to mark the contact discontinuity
($-$0.06) and the corresponding radius.  {\em Bottom}: Fe K$\alpha$
emission with continuum subtracted.  The crosses are the data and the
solid histogram is the best-fit thin shell model.  The dashed vertical
line marks the location of the inner edge of the best-fit thin shell
model and the two dotted lines show the 1-$\sigma$ allowed range.
\label{radprof}}
\end{figure}

\clearpage
\begin{figure}
\epsscale{1.0}
\plotone{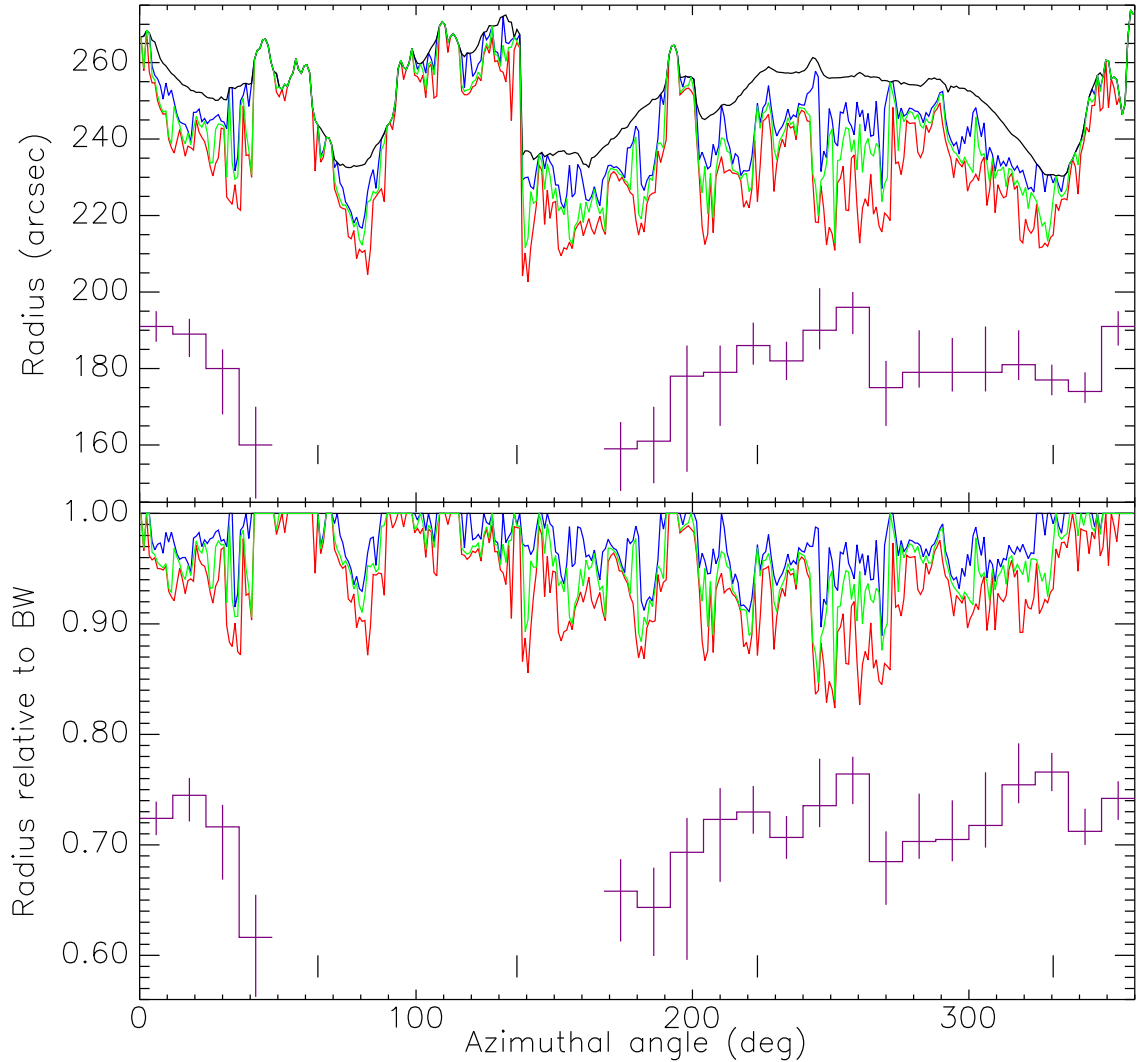}
\caption{{\em Top}: Radius of the blast wave (BW, black), contact
discontinuity (CD, 2-$\sigma$: red, 3-$\sigma$: green, 4-$\sigma$:
blue), and reverse shock (RS, purple) versus azimuthal angle for
Tycho's SNR.  The error bars on the RS are 90\%. The four small
vertical tic marks near the bottom of the figure indicate where the
gaps between the four ACIS-I CCD chips lie. North corresponds to
0$\degr$ and angles increase counterclockwise.  {\em Bottom}: As in
the top panel, with ratio of CD to BW and RS to BW plotted versus
azimuthal angle.
\label{radplot}
\label{ratio}}
\end{figure}

\clearpage
\begin{figure}
\epsscale{1.0}
\plotone{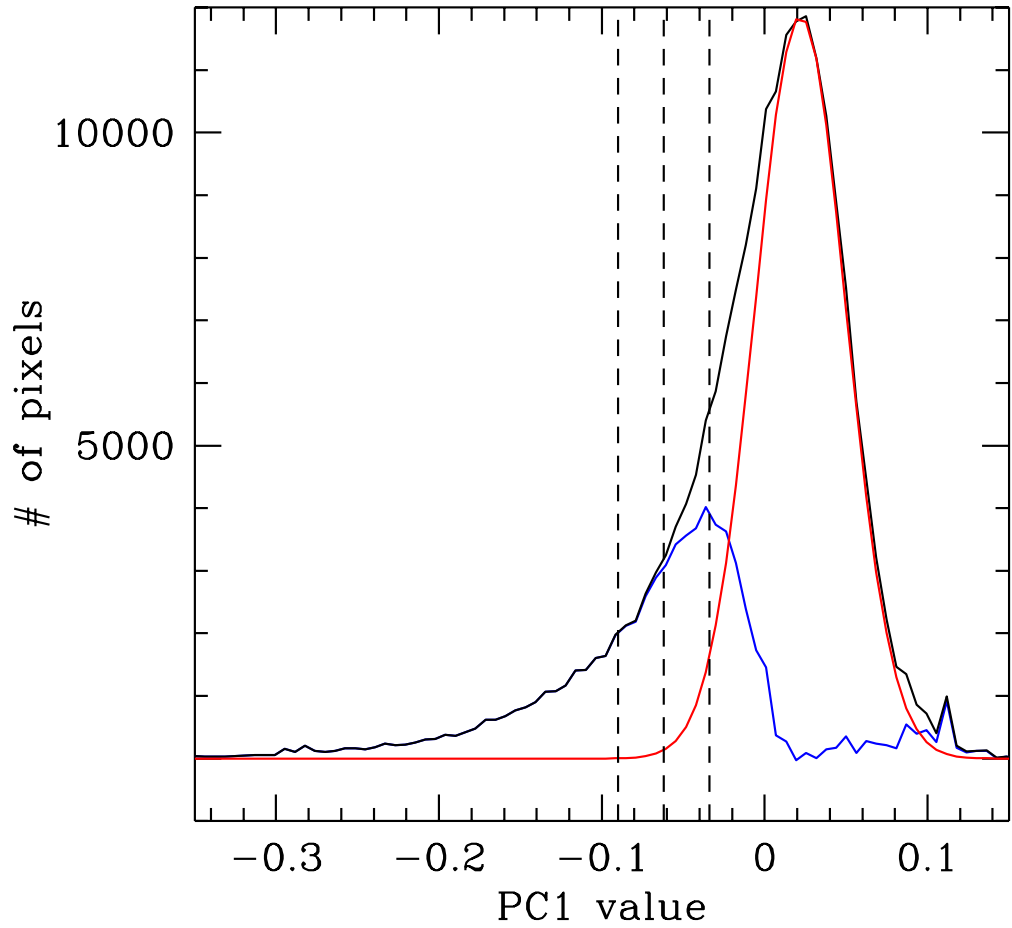}
\caption{Distribution of PC1 values (black) overlaid with best-fit
Gaussian (red, $\sigma$=0.028), which represents the pixels with
thermal spectra.  The difference between these two is shown in blue,
and the dashed lines mark PC1 values 2$\sigma$, 3$\sigma$, and
4$\sigma$ away from the peak of the Gaussian.
\label{pc1dist}}
\end{figure}

\clearpage
\begin{figure}
\epsscale{.8}
\plotone{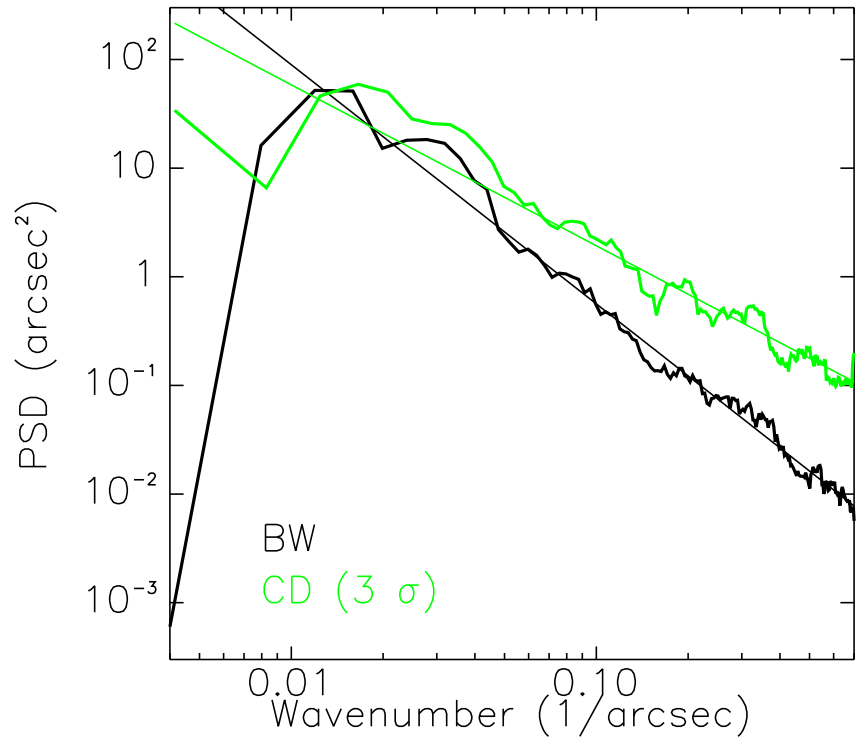}
\caption{Power spectrum of radius fluctuations from
Figure~\ref{radplot} versus wavenumber for the blast wave (BW) and
contact discontinuity (CD).  Each case is well described by a power
law as plotted.
\label{fft}}
\end{figure}

\clearpage
\begin{figure}
\epsscale{1.0}
\plotone{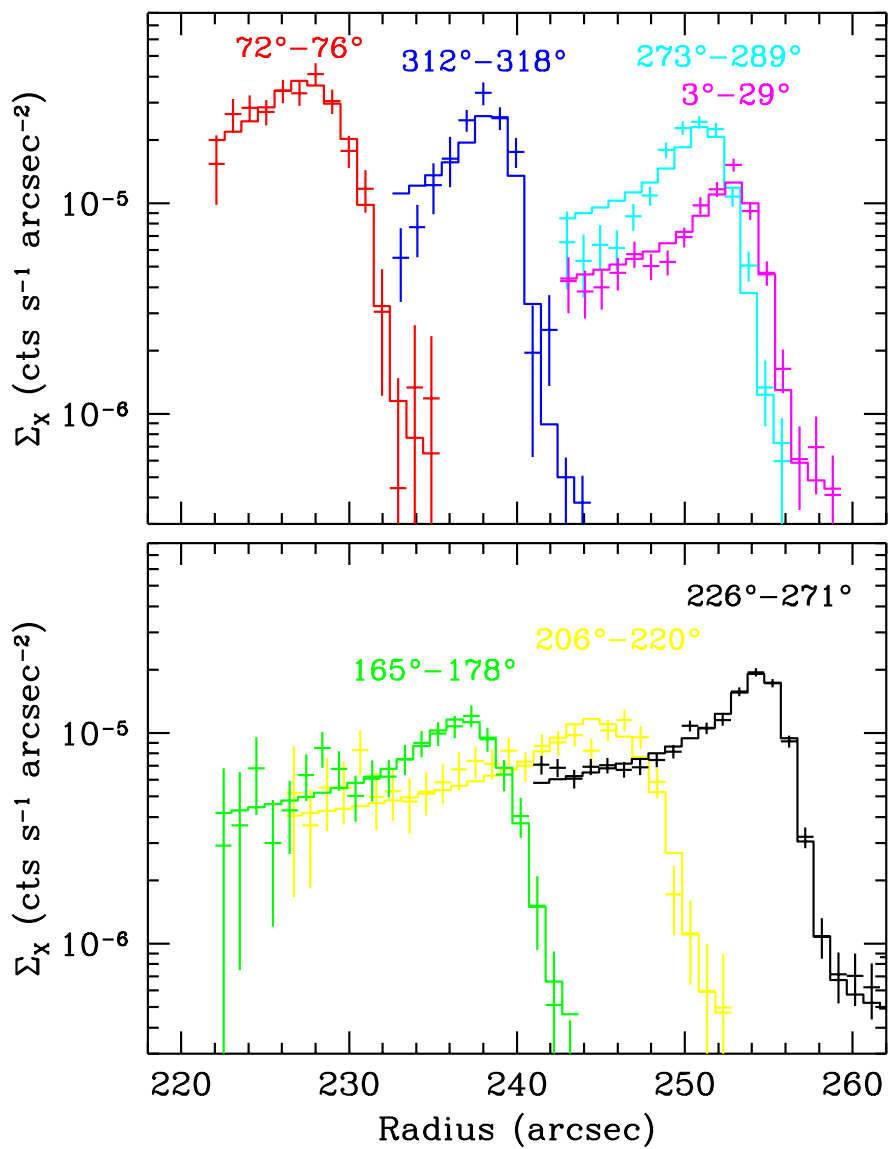}
\caption{Radial surface brightness profiles in the 4--6 keV band for
seven regions around the rim.  The histogram overlaid on each profile
is the best-fit geometrically-thin uniform-emissivity spherical shell
model.  North corresponds to 0$\degr$ and angles increase
counterclockwise.
\label{rimprof}}
\end{figure}

\clearpage
\begin{figure}
\epsscale{1.0}
\plotone{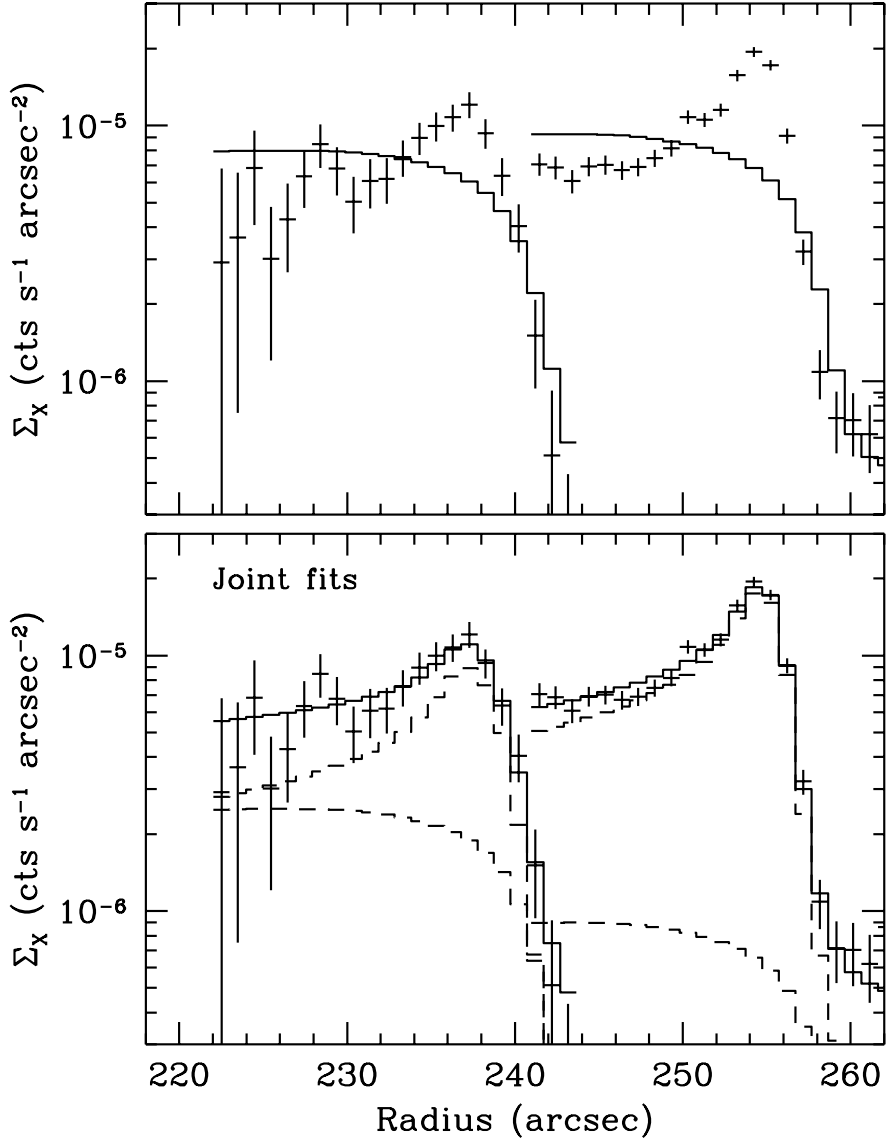}
\caption{Radial surface brightness profiles in the 4-6 keV band for
two selected regions around the rim (azimuthal angles
165$\degr$--176$\degr$ on the left and 226$\degr$--271$\degr$ on the
right).  {\em Top}: The histogram overlaid on each profile is the
best-fit model of thermal emission from shocked ambient medium based
on our 1D hydrodynamic and ionization calculations.  {\em Bottom}: The
solid histogram is the joint model of a geometrically-thin
uniform-emissivity spherical shell and the thermal shocked plasma
model. The dashed histograms show the contributions from the two cases
separately. This is not the best fit, but rather the upper limit (90\%
confidence) allowed by the data for the thermal shocked plasma model.
\label{rimprof_therm}}
\end{figure}

\clearpage
\begin{figure}
\epsscale{0.8}
\plotone{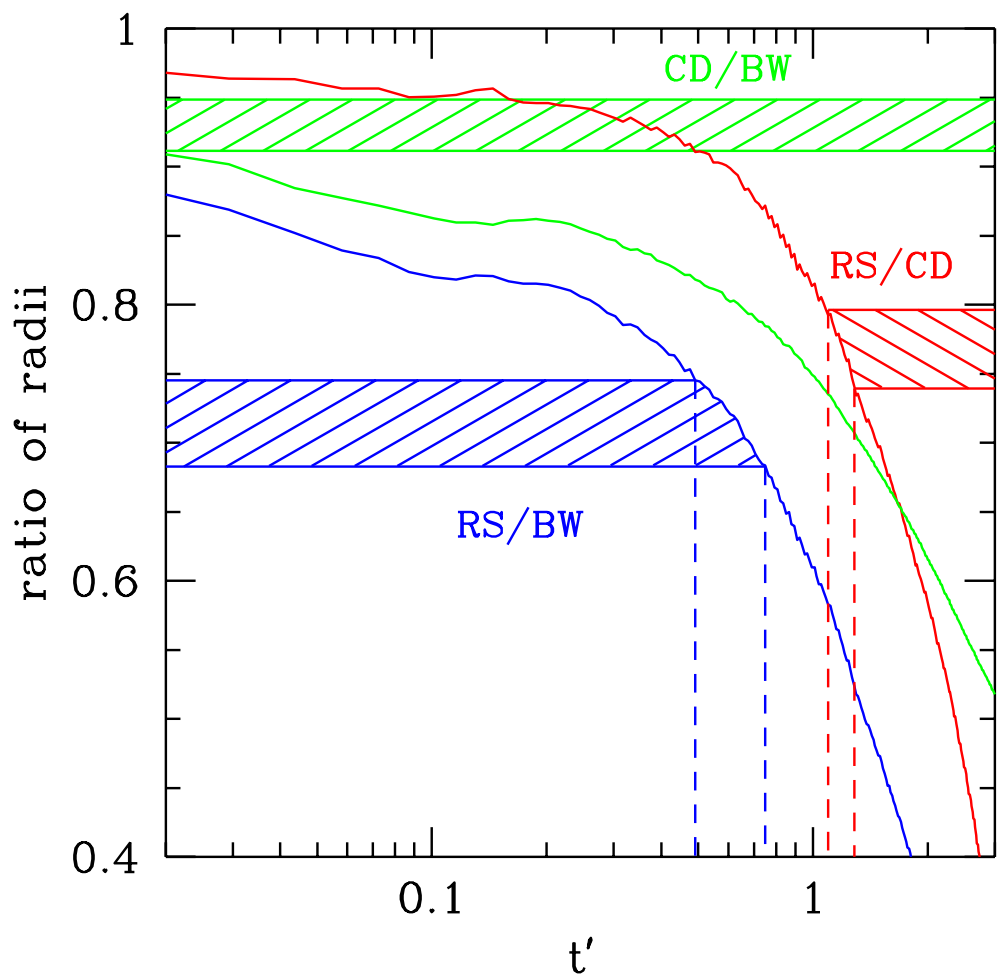}
\caption{Time evolution of the various ratios of radii for the BW, CD,
and RS as labelled.  The horizontal axis corresponds to time in
normalized units.  The curves are from a 1-D hydrodynamical
calculation \citep[model DDTc from][]{bad05} and the cross-hatched
bands show the allowed ranges from our analysis of Tycho's SNR.}
\label{ratddtc}
\end{figure}

\clearpage
\begin{figure}
\epsscale{0.8}
\plotone{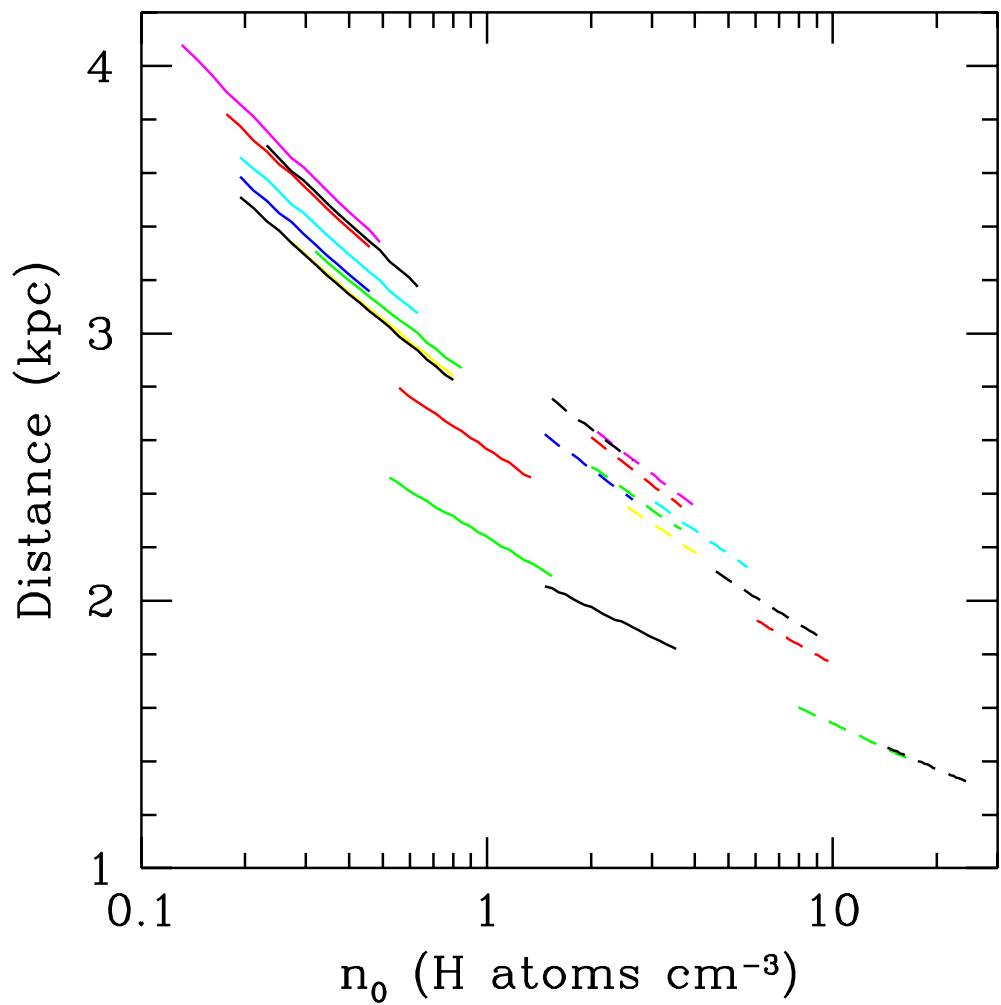}
\caption{Distance vs. ambient density estimates for Tycho's SNR from a
variety of SN Ia explosion models.  Each model provides a range of
allowed values for distance vs.~ambient density that are consistent
with the observed limits on the ratio of RS:BW radii (solid curves on
left) or the ratio of RS:CD radii (dashed curves on right).  The
different colors corresond to different models; for our purposes it is
not necessary to distinguish between them, since in no case do the
allowed ranges overlap.
\label{distden}}
\end{figure}

\end{document}